\documentclass[usenatbib,useAMS]{mn2e}
\usepackage[fleqn,tbtags]{amsmath} 
\usepackage{aas_macros,amssymb}
\usepackage{txfonts}
\usepackage{natbib,graphicx}
\pubyear{2011}
\title[Multiple Black Holes]{Formation of galactic nuclei with
  multiple supermassive black holes at high redshifts}
\author[Kulkarni \& Loeb]{Girish Kulkarni$^{1,2}$\thanks{Email:
    girish@hri.res.in}, Abraham Loeb$^{1}$\thanks{Email:
    aloeb@cfa.harvard.edu}\\ $^1$Institute for Theory \& Computation,
  Harvard University, 60 Garden Street, Cambridge, MA 02138
  USA\\ $^2$Harish-Chandra Research Institute, Chhatnag Road, Jhunsi,
  Allahabad 211019 India } \date{}

\begin{document}
\maketitle 

\begin{abstract}
We examine the formation of groups of multiple supermassive black
holes (SMBHs) in gas-poor galactic nuclei due to the high merger rate
of galaxies at high redshifts.  We calculate the relative likelihood
of binary, triple, and quadruple SMBH systems, by considering the
timescales for relevant processes and combining merger trees with
N-body simulations for the dynamics of stars and SMBHs in galactic
nuclei.  Typical haloes today with mass $M_0\approx 10^{14}$ M$_\odot$
have an average mass $M_{z=6}=5\times 10^{11}$ M$_\odot$ at $z\sim 6$,
while rare haloes with current mass $M_0\gtrsim 10^{15}$ M$_\odot$
have an average mass $M_{z=6}=5\times 10^{12}$ M$_\odot$ at that
redshift.  These cluster-size haloes are expected to host single
galaxies at $z\sim 6$.  We expect about 30\% galaxies within haloes
with present-day mass $M_0\approx 10^{14}$ M$_\odot$ to contain more
than two SMBHs at redshifts $2\lesssim z\lesssim 6$.  For larger
present-day haloes, with $M_0\gtrsim 10^{15}$ M$_\odot$, this fraction
is almost 60\%.  The existence of multiple SMBHs at high redshifts can
potentially explain the mass deficiencies observed in the cores of
massive elliptical galaxies, which are up to $5$ times the mass of
their central BHs.  Multiple SMBHs would also lead to an enhanced rate
of tidal disruption of stars, modified gravitational wave signals
compared to isolated BH binaries, and slingshot ejection of SMBHs from
galaxies at high speeds in excess of $2000$ km s$^{-1}$.

\end{abstract}


\section{Introduction} 
\label{sec:intro}

Most local galaxies host supermassive black holes (SMBHs) at their
centres \citep{1998Natur.395A..14R, 2005SSRv..116..523F}.  The SMBH
mass $M_{\rm bh}$ is correlated with properties of the spheroidal
nucleus of the host galaxy, such as velocity dispersion
\citep{2000ApJ...539L...9F, 2000ApJ...539L..13G, 2002ApJ...578...90F,
  2009ApJ...698..198G} and luminosity \citep{1998AJ....115.2285M,
  2002MNRAS.331..795M, 2003ApJ...589L..21M, 2009ApJ...698..198G}.
Detection of bright quasars at redshifts $z\gtrsim 6$
\citep{2001AJ....122.2833F, 2011Natur.474..616M} suggests that SMBHs
with masses as high as $\sim 2\times 10^9$ M$_\odot$ already existed
at $z\sim 7$.  In the standard $\Lambda$CDM cosmological model, growth
of galaxies is hierarchical and galaxy mergers are expected to be
particularly frequent at redshifts $z\sim 6$--$20$.  As galaxies
merge, their central SMBHs can grow through coalescence and accretion
of gas.  It is commonly postulated that SMBHs at lower redshifts grew
out of seed black holes (BHs) in the first galaxies
\citep{1994ApJ...432...52L, 1995ApJ...443...11E, 2000MNRAS.311..576K,
  2001ApJ...558..535M, 2003ApJ...596...34B, 2003ApJ...582..559V,
  2006ApJS..163....1H, 2009ApJ...696.1798T}.

Existing merger tree models are based on the assumption that any
binary black hole system, which inevitably forms in a galaxy's merger
history, coalesces on a short time-scale.  However, the evolution of
SMBH binaries is a complex open problem and it is unclear if a binary
can merge within a Hubble time \citep {2005LRR.....8....8M}.  One
expects that during a merger event of two galaxies, the dynamics of
their constituents SMBHs would proceed in three stages
\citep{1980Natur.287..307B}.  In the first stage, the SMBHs sink to
the centre of the gravitational potential of the merger remnant by
dynamical friction and form a gravitationally bound binary.  The
newly-formed binary continues to lose energy and angular momentum
through its global gravitational interaction with many stars until the
separation between the SMBHs reduces to a value at which the dominant
mechanism of energy loss is the 3-body interaction between the binary
and individual stars.  This is the second stage of the binary's
evolution, and is known as the `hard stage.'  The precise definition
of a hard SMBH binary varies in the literature, but it is commonly
assumed that the binary becomes hard when its semi-major axis $a$
reaches a value given by \citep{2002MNRAS.331..935Y}
\begin{equation}
a\approx a_h\equiv\frac{Gm}{4\sigma^2}=2.8\left(\frac{m}{10^8
  \mathrm{M}_\odot}\right)\left(\frac{200\;\mathrm{km}\,\mathrm{s}^{-1}}{\sigma}\right)^2
\mathrm{pc},
\label{eqn:ahard}
\end{equation}
where stars in the galactic nucleus are assumed to have a
one-dimensional velocity dispersion $\sigma$, and $m$ denotes the mass
of the lighter SMBH.  Finally, once the SMBH separation decreases to a
small-enough value, gravitational wave emission becomes the dominant
mode of energy loss and the SMBHs coalesce rapidly.  This is the third
stage of the SMBH binary evolution.  The value of semi-major axis $a$
at which the coalescence time scale due to gravitational wave emission
alone is $t$ is given by \citep{peters1964, 2010PhRvD..81d7503L}
\begin{equation}
a(t)\equiv a_\mathrm{gw}(t)=4.3\times
10^{-3}\left(\frac{t}{10^{5}\mathrm{yr}}\right)^{1/4}\left(\frac{M}{2\times
  10^{8}\mathrm{M}_\odot}\right)^{3/4}\!\mathrm{pc},
\label{eqn:agw}
\end{equation}
where $M$ is the total mass of the binary, and we have considered two
SMBHs with mass $10^8$ M$_\odot$ each on a circular orbit (with
shorter time scale at increasing eccentricity).  Gravitational wave
emission takes over as the dominant mode of energy loss when
$a=a_\mathrm{gw}(t_h)$, where $t_h$ is the hardening time scale.

Among these three stages of evolution of an SMBH binary, the largest
uncertainty in the binary's lifetime originates from the hard stage,
which can be the slowest of the three stages since the binary quickly
ejects all low angular momentum stars in its vicinity, thus cutting
off its supply of stars.  This is known as the ``final parsec
problem'' \citep{2003AIPC..686..201M}.  For example,
\citet{2002MNRAS.331..935Y} studied coalescence of SMBH binaries in a
sample of galaxies observed by \citet{1997AJ....114.1771F} and found
that spherical, axisymmetric or weakly triaxial galaxies can all have
long-lived binary SMBHs that fail to coalesce.  Similarly,
\citet{2005LRR.....8....8M} found that the time spent by a binary is
less than $10^{10}$ yr only for binaries with very low mass ratios
($\lesssim 10^{-3}$).\footnote{However, for such low mass ratios the
time taken by the lighter black hole to reach the galactic nucleus due
to dynamical friction is itself expected to exceed the Hubble time.}
Furthermore, \citet{2005LRR.....8....8M} showed that a binary may not
be able to interact with all the stars in its loss cone, thereby
increasing the time spent in the hard stage even further; they found
that in a nucleus with a singular isothermal sphere stellar density
profile, an equal-mass binary will stall at a separation of $a\approx
a_h/2.5$, where we have defined $a_h$ in Equation (\ref{eqn:ahard}).
The final separation is expected to be even higher for galaxies with
shallower density profiles.

Several ways have been discussed in the literature to efficiently
extract energy and angular momentum from a hard SMBH binary and
overcome the final parsec problem.  An example is the work by
\citet{2002ApJ...567L...9A}, who suggested that gas can catalyse the
coalescence of a hard SMBH binary by serving as an effective sink for
the binary's angular momentum.  In particular, they found that a
binary with a separation of $0.1$ pc embedded in a gaseous accretion
disk would merge in $10^7$ years without significant enhancement in
the gas accretion rate.  Similarly, \citet{2004ApJ...607..765E,
  2005ApJ...630..152E} found that in SPH simulations, clouds of hot
gas ($T_\mathrm{gas}\approx T_\mathrm{virial}$) can induce decay of
orbits of embedded binary point masses due to gravitational drag.  A
caveat to these studies is that feedback from gas accretion onto the
SMBHs can remove the rest of the gas from the merger remnant before
the binary coalesces.  However, stellar dynamical processes could also
accelerate binary coalescence, without gas.  For example,
\citet{2004ApJ...606..788M} considered the effect of chaotic orbits in
steep triaxial potentials. They found that stars are supplied to the
central black hole at a rate proportional to the fifth power of the
stellar velocity dispersion and that the decay rate of a central black
hole binary would be enhanced even if only a few percent of the stars
are on chaotic orbits, thus solving the final parsec problem.  As
another example, it was suggested that a third SMBH closely
interacting with a hard SMBH binary can reduce the binary separation
to a small value either due to the eccentricity oscillations induced
in the binary via the Kozai-Lidov mechanism
\citep{2002ApJ...578..775B} or due to repopulation of the binary's
loss cone due to the perturbation in the large-scale potential caused
by the third black hole \citep{2007MNRAS.377..957H}.
\citet{2002ApJ...578..775B} found that the merger time scale of an
inner circular binary can be shortened by as much as an order of
magnitude, and that general relativistic precession does not destroy
the Kozai-Lidov effect for hierarchical triples that are compact
enough.

In summary, there is substantial uncertainty in the current
understanding of the evolution of binary SMBHs.  Clearly, if the SMBH
binary coalescence time is longer than the typical time between
successive major mergers of the galaxy, then more than two SMBHs may
exist in the nucleus of a merger remnant.  We study this possibility
in this paper.  We calculate the relative likelihood of binary,
triple, and quadruple SMBH systems, by considering the timescales for
relevant processes and combining galaxy merger trees with
direct-summation N-body simulations for the dynamics of stars and
SMBHs in galactic nuclei.  An obvious question regarding galactic
nuclei with multiple SMBHs is whether such systems can be long-lived.
We consider this question here.  Finally, systems with multiple SMBHs
are likely to be interesting because of observational effects
involving their effect on the properties of the host bulge, the
enhancement in the rate of tidal disruption of stars, their associated
gravitational wave and electromagnetic signals, and the slingshot
ejection of SMBHs at high speeds.  We study some of these effects.

In \S \ref{prev} we review previous results on galactic nuclei with
more than two SMBHs.  We present simple analytical arguments regarding
the formation and evolution of such systems in \S
\ref{sec:formation_analytic} and \ref{sec:evolution_analytic}.
Details about our numerical simulations are described in \S
\ref{sec:simulations}, with their results shown in \S
\ref{sec:results_numerical}.  We consider the observational signatures
of our findings in \S \ref{sec:signatures}.  Finally, we discuss and
summarise our primary findings in \S \ref{sec:conclusions}.

\section{Previous work}
\label{prev}

Galactic nuclei with multiple SMBHs were first studied by
\citet{1974ApJ...190..253S}, who computed orbits of three and four
SMBH systems by sampling the parameter space of the problem.  They
showed that if an infalling SMBH is lighter than the components of the
pre-existing binary, then the most probable outcome is a slingshot
ejection in which the infalling SMBH escapes at a velocity that is
about a third of the orbital velocity of the binary.
\citet{1976A&A....46..435V} further showed that the ejection velocity
can be significantly enhanced if drag forces due to gravitational
radiation are accounted for in the three-body dynamics.  The formation
of systems with three or four SMBHs in a hierarchical merger of smooth
galactic potentials was first studied by \citet{1990ApJ...348..412M}
and \citet{1994ApJS...95...69V} with the objective of understanding
the structure of extragalactic radio sources.  This line of work was
extended to binary-binary scattering of SMBHs by
\citet{2001A&A...371..795H}, and by \citet{2007MNRAS.377..957H}, who
studied repeated triple interactions in galactic nuclei.  Both of
these studies used cosmologically consistent initial conditions based
on the extended Press-Schechter theory.  Systems with a larger number
of black holes were studied by \citet{1992MNRAS.259P..27H} and
\citet{1994ApJ...437..184X} using simple analytical models and
numerical calculations of massive particles in smooth galactic
potentials.  \citet{1994ApJ...437..184X} concluded that the
most-likely outcome in these cases is one in which most black holes
are ejected and the galactic center is left with zero, or one, or two
black holes.  Finally, full N-body simulations of galactic nuclei with
constituent SMBHs were performed for the case of two successive
mergers by \citet{1996ApJ...465..527M}, \citet{1997ApJ...478...58M},
and \citet{2006ApJ...651.1059I}.  Much of this work on SMBHs was based
on earlier studies of stellar-mass black holes in globular clusters.
\citet{1993Natur.364..423S} and \citet{1993Natur.364..421K} considered
the evolution of $\sim 100$ stellar mass black holes in globular
clusters.  They concluded that after mass segregation, most of these
black holes are ejected out on a short time scale, and the globular
cluster is left with none or a few black holes.  Mass segregation and
associated effects of stellar-mass black holes in a galactic nucleus
with a central SMBH was also considered \citep{2000ApJ...545..847M,
  2006ApJ...649...91F}.

The possible formation of systems with multiple SMBHs due to
successive galactic mergers arises naturally in any model describing
the hierarchical assembly of galaxies.  One approach to modeling SMBH
growth involves constructing semi-analytic prescriptions of various
characteristic processes, like mergers of galaxies, formation of
spheroids, star formation, and gas thermodynamics, coupled with merger
trees of dark matter haloes.  This approach has been adopted, for
example, by \citet{2000MNRAS.311..576K}, who also extended it to study
possible formation of multiple SMBH systems and implications for the
$M_{\rm bh}$--$\sigma$ relation and density profiles observed in
luminous elliptical galaxies \citep{2002MNRAS.336L..61H}.  Another
study by \citet{2003ApJ...582..559V} followed merger trees of dark
matter haloes and their component SMBHs using Monte Carlo realizations
of hierarchical structure formation in the $\Lambda$CDM cosmology.
They modeled dark matter haloes as singular isothermal spheres and
calculated the inspiral of less massive halos in more massive ones by
using the Chandrasekhar formula for dynamical friction.  Gas accretion
to the SMBHs was modeled so as to reproduce the empirical $M_{\rm
  bh}$--$\sigma$ relation and the SMBH dynamics was described with
analytic prescriptions.  In particular, the coalescence time of hard
SMBH binaries was calculated from a set of coupled differential
equations based on scattering experiments involving the ejection of
stellar mass from the loss cone due to the hard SMBH binary and the
resultant change in the hardening rate \citep{1996NewA....1...35Q,
  2000ASPC..197..221M}.  For galaxies that underwent another major
merger before their constituent binary SMBH coalesced, a three-body
interaction was implemented between the binary and the intruder SMBH.
They found that the smallest SMBH was kicked out of the galaxy in 99\%
of cases, while the binary escapes the galaxy in 8 \% of cases.  Thus,
a significant fraction of galactic nuclei could end up with no SMBHs
or offset SMBHs with mass lower than that expected from the $M_{\rm
  bh}$--$\sigma$ relation.  These results were later extended to
incorporate recoil in the SMBH merger remnant due to asymmetric
emission of gravitational waves, which mainly affected the $M_{\rm
  bh}$--$\sigma$ relation for low-mass haloes by increasing the
scatter \citep{2006ApJ...650..669V, 2007ApJ...663L...5V,
  2011MNRAS.412.2154B}.  Similar semi-analytic models were studied by
several other authors to understand the assembly of $z\sim 6$ quasars.
However, most of these models ignored the dynamics of multiple SMBHs
and assumed prompt coalescence \citep{1999ApJ...521L...9H,
  2003ApJ...595..614W, 2004ApJ...614L..25Y, 2009ApJ...696.1798T,
  2009ApJ...704...89S}.  As a result, these models did not treat
systems with multiple SMBHs.

Lastly, SMBH assembly has also been studied using smooth particle
hydrodynamic simulations that attempted to calculate effects of both
the gas physics as well as the gravitational dynamics of the
large-scale structure within and around galaxies
\citep{2006ApJS..163....1H, 2007MNRAS.380..877S, 2007ApJ...665..187L,
  2007ApJ...669...45H}.  However, due to poor mass resolution and
particle smoothening, these simulations cannot accurately calculate
the detailed dynamics of a multiple SMBH systems.  Indeed, in most of
these studies, black hole coalescence occurs on scales smaller than
the smoothening length, which is usually much larger than the expected
separation of a hard SMBH binary.  As a result, SMBH coalescence is
implemented via a subgrid model. Here, we explore for the first time
numerical simulations that incorporate the cosmological process of
galaxy mergers in the cosmological context along with an accurate
treatment of black hole dynamics.

\section{Formation of multiple-SMBH systems}
\label{sec:formation_analytic}

Unless they coalesce rapidly, or get kicked out of the host galactic
nucleus, we expect multi-SMBH systems to form in galactic nuclei at
high redshift due to mergers of galaxies if the typical black hole
coalescence timescale is longer than the feeding timescale of new
incoming black holes.  In this section, we establish a simple
theoretical framework for this formation path using analytical
estimates of its relevant timescales: {\it (i)} the major merger time
scale of galaxies; {\it (ii)} the time scale on which a satellite
galaxy sinks to the center of a host galaxy so that a close
interaction between SMBHs can occur; and {\it (iii)} the time scale of
SMBH coalescence.

\subsection{Time scale of incoming SMBHs}

\begin{figure}
\begin{center}
\includegraphics[scale=0.6]{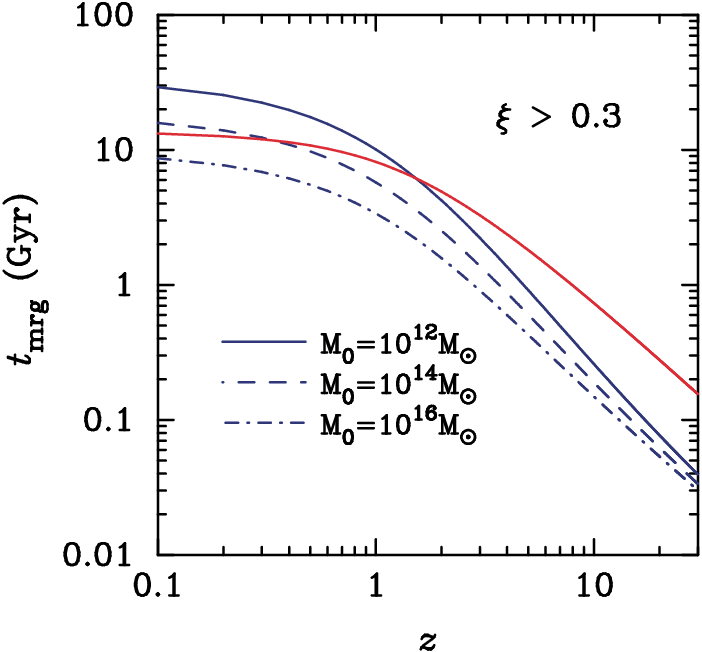} 
\end{center}
\caption{Halo major merger time scale (mass ratio $>0.3$), according
  to equation (\ref{eqn:tmrg}), for haloes with mass $M_0=10^{12}$
  M$_\odot$ (blue solid line), $10^{14}$ M$_\odot$ (blue dashed line)
  and $10^{16}$ M$_\odot$ (blue dot-dashed line).  The Hubble time is
  shown by the solid red curve.  Major mergers are more frequent at
  higher redshifts.  On average, Milky Way-sized haloes are not
  expected to undergo a major merger for $z\lesssim 1$.  Galaxy major
  merger time scale is always longer than the subsequent dynamical
  friction time scale.}
\label{fig:tmrg}
\end{figure}

\citet{2010MNRAS.406.2267F} have quantified the average merger rate of
dark matter haloes per halo per unit redshift per unit mass ratio for
a wide range of halo mass, progenitor mass ratios and redshift.  The
result is given by a fitting formula derived from the Millennium
\citep{2005Natur.435..629S} and Millennium-II
\citep{2009MNRAS.398.1150B} simulations:
\begin{equation}
\frac{dN}{d\xi dz}(M,\xi,z) =
A\left(\frac{M}{10^{12}\mathrm{M}_\odot}\right)^\alpha\xi^\beta\exp\left[\left(\frac{\xi}{\tilde\xi}\right)^\gamma\right](1+z)^\eta.
\label{eqn:mrgrate}
\end{equation}
Here, $M$ is the halo mass at redshift $z$, and $\xi$ is the mass
ratio of progenitors.  Mergers with $\xi > 0.3$ are considered major
mergers.  The best fit values of various parameters are
$\alpha=0.133$, $\beta=-1.995$, $\gamma=0.263$, $\eta=0.0993$,
$A=0.0104$ and $\tilde\xi=9.72\times 10^{-3}$.  The average major
merger rate per unit time is then given by
\begin{equation}
\frac{dN_m}{dt}(M,z)=\int_{0.3}^{1}d\xi\frac{dN}{d\xi
  dz}(M,\xi,z)\frac{dz}{dt}.
\label{eqn:majormrgrate_time}
\end{equation}
\citet{2010MNRAS.406.2267F} also provide a fitting formula for average
mass growth rate of halos that can be used to calculate the halo mass
at redshift $z$ for use in equation (\ref{eqn:mrgrate}),
\begin{equation}
\dot M(z) =
46.1\,\frac{\mathrm{M}_\odot}{\mathrm{yr}}\,(1+1.1z)\sqrt{\Omega_m(1+z)^3+\Omega_\Lambda}\,\left(\frac{M}{10^{12}\mathrm{M}_\odot}\right)^{1.1}.
\label{eqn:halo_assembly}
\end{equation}
Using equation (\ref{eqn:majormrgrate_time}) we can now define the
time scale of major mergers for a halo as
\begin{equation}
t_\mathrm{mrg}=\left[\frac{dN_m}{dt}\right]^{-1}.
\label{eqn:tmrg}
\end{equation}
The behavior of this quantity is shown in Figure \ref{fig:tmrg} for
three halo masses that discussed here: a Milky Way-like halo that has
a mass $M_0=10^{12}$ M$_\odot$ at $z=0$, the typical halo today that
has mass $M_0=10^{14}$ M$_\odot$ at $z=0$, and rare haloes with mass
$M_0=10^{16}$ M$_\odot$ at $z=0$.  (In this paper, $M_0$ always
denotes the halo mass at redshift $z=0$.  We also refer to the average
mass of such haloes at other redshifts, by e.~g.\ $M_{z=4}$ and
$M_{z=6}$.  A halo with $M_0=10^{12}$ M$_\odot$ will have
$M_{z=6}=2\times 10^{10}$ M$_\odot$.  A halo with $M_0=10^{14}$
M$_\odot$ will have $M_{z=6}=5\times 10^{11}$ M$_\odot$.  A
cluster-size halo, with mass $M_0=10^{15}$ M$_\odot$ will have
$M_{z=6}=5\times 10^{12}$ M$_\odot$ and is expected to hold a single
galaxy at that redshift.)  This is the time scale at which we expect
new (satellite) haloes to enter the halo.  As expected, halo mergers
are more frequent at higher redshift.  At redshift $z\lesssim 1$ the
major merger time scale for a Milky Way-like halo is greater than the
Hubble time.

After two dark matter haloes have merged, the smaller halo becomes a
satellite halo within the virial radius of the host halo.  It then
takes this satellite a dynamical friction time to sink to the center
of the host halo, so that the constituent galaxies can merge.  As a
result, the timescale for major mergers of galaxies is expected to be
different that the time scale for major mergers of dark matter haloes
calculated in Equation (\ref{eqn:tmrg}).

The dynamical friction time scale is often estimated using
Chandrasekhar's formula \citep{1943ApJ....97..255C,
1993MNRAS.262..627L, 2008gady.book.....B}:
\begin{equation}
t_\mathrm{df}=\frac{f_\mathrm{df}\Theta_\mathrm{orb}}{\ln\Lambda}\frac{M_\mathrm{host}}{M_\mathrm{sat}}t_\mathrm{dyn},
\label{tdf_chandra}
\end{equation}
where $M_\mathrm{host}$ and $M_\mathrm{sat}$ are the masses for the
host and satellite haloes respectively, $\ln\Lambda$ is the coulomb
logarithm, $\Theta_\mathrm{orb}$ is a function of the orbital energy
and angular momentum of the satellite, $f_\mathrm{df}$ is an
adjustable parameter of order unity and $t_\mathrm{dyn}$ is the halo
dynamical time scale calculated at the virial radius.  Equation
(\ref{tdf_chandra}) is valid only in the limit of small satellite mass
in an infinite, isotropic and homogeneous collisionless medium.
Still, it has been used in the literature even for large satellite
masses by modifying the Coulomb logarithm.  In recent years,
deviations from predictions by equation (\ref{tdf_chandra}) have been
reported in both the $M_\mathrm{sat}\ll M_\mathrm{host}$ and
$M_\mathrm{sat}\lesssim M_\mathrm{host}$ regimes
\citep{2003MNRAS.341..434T, 2007MNRAS.375.1189M, 2008MNRAS.383...93B,
2008ApJ...675.1095J, 2009MNRAS.395.1376W}.

To correct the problems associated with Chandrasekhar's formula,
several groups have developed full dynamical models of evolution of
merging haloes \citep{2001ApJ...559..716T, 2003ApJ...589..752G,
2003MNRAS.341..434T, 2005ApJ...624..505Z}.  For example, one of the
approaches to overcome the limits of Chandrasekhar's formula is the
theory of linear response (TLR; \citealt{1999ApJ...525..720C}).  TLR
captures the backreaction of the stellar distribution to the intruding
satellite by correlating the instantaneous drag force on it with the
drag force at an earlier time via the fluctuation-dissipation theorem.
Tidal stripping of a satellite halo is an important ingredient in this
formulation.  In a singular isothermal sphere with 1D velocity
dispersion $\sigma$ and density profile $\rho(r)=\sigma^2/[2\pi
Gr^2]$, TLR predicts a sinking time
\begin{equation}
t_\mathrm{df}=1.17\frac{r^2_\mathrm{cir}V_\mathrm{cir}}{GM_\mathrm{sat}\ln\Lambda}\epsilon^\alpha,
\label{tdf_tlr}
\end{equation}
where $\epsilon$ is the circularity (defined as the ratio between the
angular momentum of the current orbit relative to that of a circular
orbit of equal energy), $r_\mathrm{cir}$ and $V_\mathrm{cir}$ are the
initial radius and velocity of the circular orbit with the same energy
of the actual orbit, and $M_\mathrm{S}$ is the mass of the incoming
satellite halo.  Numerical simulations suggest a value of $0.4-0.5$
for the exponent $\alpha$ \citep{1999ApJ...515...50V,
1999ApJ...525..720C, 2003ApJ...582..559V}.

Given the limitations of analytical treatments, we turn to results of
numerical simulations to understand the dynamical friction time scale.
Using N-body simulations, \citet{2008MNRAS.383...93B} give a fitting
formula that accurately predicts the time-scale for an extended
satellite to sink from the virial radius of a host halo down to the
halo's centre for a wide range of mass ratios and orbits (including
a central bulge in each galaxy changes the merging time scale by
$\lesssim 10$ \%).  Their fitting formula is given by
\begin{equation}
\frac{t_\mathrm{df}}{t_\mathrm{dyn}}=A\frac{\xi^{-b}}{\ln(1+1/\xi)}\exp\left[c\frac{j}{j_\mathrm{cir}(E)}\right]\left[\frac{r_\mathrm{cir}(E)}{r_\mathrm{vir}}\right]^d,
\label{tdf_fit}
\end{equation}
where $A=0.216$, $b=1.3$, $c=1.9$ and $d=1.0$.  Here $\xi$ is the mass
ratio $M_\mathrm{sat}/M_\mathrm{host}$, $j$ is the specific angular
momentum of the satellite halo, and $j_\mathrm{cir}$ is the specific
angular momentum of a circular orbit with the same energy $E$.  This
formula is expected to be valid for $0.025\leq\xi\leq 1.0$, and for
circularities $0.3\leq\eta\equiv j/j_\mathrm{cir}(E)\leq 1.0$.  Most
likely value of circularity in dark matter simulations is $\eta\approx
0.5$ \citep{2005MNRAS.358..551B, 2005ApJ...624..505Z,
  2006A&A...445..403K}.  Lastly, it is valid for range of orbital
energy $-0.65\leq r_\mathrm{cir}(E)/r_\mathrm{vir}\leq 1.0$.  This
covers the peak value of distribution seen in cosmological N-body
simulations.  We fix $r_\mathrm{cir}(E)/r_\mathrm{vir}=1.0$ and
$\eta=0.5$, which are the typical values found in simulations.

\begin{figure}
\begin{center}
\includegraphics[scale=0.6]{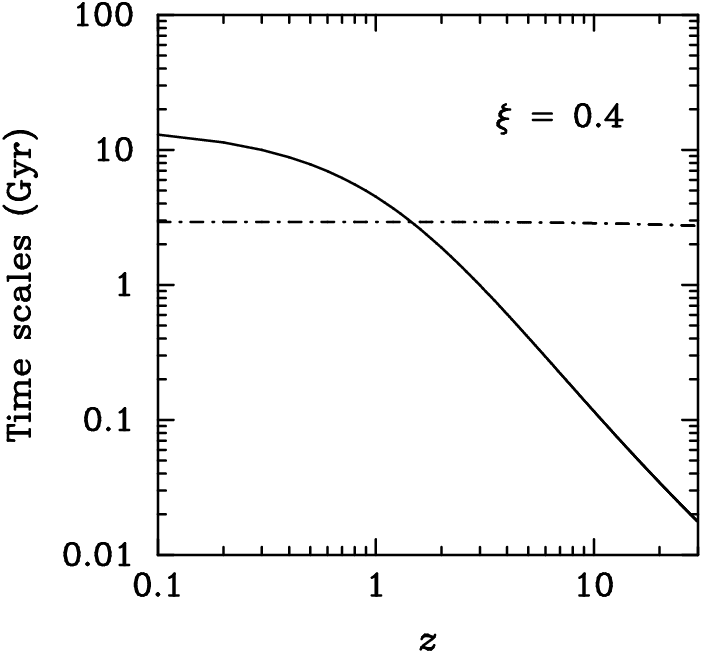} 
\end{center}
\caption{A comparison between the feeding time scale of incoming black
  holes $t_\mathrm{in}$ (black solid line; Eq. \ref{eqn:tin}) and the
  time scale of black hole coalescence $t_\mathrm{coal}$ (black
  dot-dashed line; Eq. \ref{eqn:tcoal}), for a halo mass $M_0=10^{12}$
  M$_\odot$ and considering only mergers with a mass ratio $\xi=0.4$.
  The coalescence time $t_\mathrm{coal}$ has only a weak dependence on
  redshift because its dependence on $M_\mathrm{bh}$ and $\sigma$
  cancel out due to the $M_{\rm bh}$--$\sigma$ relation.  This figure
  shows that at high redshift new black holes would arrive to the
  center of a galaxy faster than they could merge via dynamical
  processes.}
\label{fig:tcoal}
\end{figure}

We can now obtain the instantaneous merger rate of galaxies by combining the halo merger rate and dynamical friction time scale.
We closely follow the method of \citet{2009ApJ...704...89S} and write
\begin{equation}
B_\mathrm{gal}(M,\xi,z)=B[M,\xi,z_e(z,\xi)]\frac{dz_e}{dz},
\end{equation}
where $B(M,\xi,z)$ (per unit volume per unit mass per unit redshift
per unit mass ratio) is the instantaneous merger rate of halos with
mass $M$, progenitors with mass ratio $\xi$ at redshift $z$,
$B_\mathrm{gal}$ is the same quantity for galaxies.  The redshift
$z_e(z,\xi)$ is a function of $z$ and $\xi$, and is given implicitly
by
\begin{equation}
t(z)-t(z_e)=t_\mathrm{mrg}(\xi,z_e),
\end{equation}
where $t(z)$ is the cosmic time at redshift $z$.
\citet{2009ApJ...704...89S} finds that $dz_e/dz$ is almost constant at
all redshifts for $\xi=0.1-1$ and can be approximated by
\begin{equation}
\frac{dz_e}{dz}\approx1+0.09[\xi^{1.3}\ln(1+1/\xi)]^{-1},
\end{equation}
for the fitting formula in equation (\ref{tdf_fit}).  We assume this
form in our calculations.  Once we have calculated
$B_\mathrm{gal}(M,\xi,z)$, we normalize it by $n(M,z)$, the abundance
of haloes of mass $M$ at redshift $z$.  We use the Sheth-Tormen mass
function \citep{1999MNRAS.308..119S} to calculate $n(M,z)$.  This
gives us the galaxy merger rate \emph{per halo} per unit $\xi$ per
unit redshift, which is the galaxy's counterpart of equation
(\ref{eqn:mrgrate}), and which we denote by $dN_\mathrm{gal}/dz$.  The
rate of mergers of galaxies is the rate at which new black holes are
added to the host halo's nucleus.  Thus, the time scale of incoming
black holes is
\begin{equation}
t_\mathrm{in}=\left[\frac{dN_\mathrm{gal}}{dz}\frac{dz}{dt}\right]^{-1}.
\label{eqn:tin}
\end{equation}
The result is shown by the solid black line in Figure \ref{fig:tcoal}
for a mass ratio of $\xi=0.4$ and a halo that has mass of $10^{12}$
M$_\odot$ at $z=0$.

\subsection{Binary SMBH coalescence time scale}

In order to find whether there is a generic possibility of formation
of systems with multiple SMBHs, we compare the time scale on which new
black holes are added to the galactic nucleus at a certain redshift
with the coalescence time scale of a binary SMBH at that redshift.

As described in \S \ref{sec:intro}, the formation and coalescence of a
black hole binary is expected to take place in three stages.  We
define the coalescence time as the time that the binary spends in the
second of these stages, that is the time from when the binary
separation is $a=a_h$, defined in equation (\ref{eqn:ahard}), up to
when the separation is $a=a_\mathrm{gr}$ at which point the binary
enters the third stage of evolution, and gravitational waves become
the dominant mechanism of energy loss.  For a hard binary, the
dominant channel through which energy is lost is three-body
interactions in which stars passing in close proximity to the binary
are ejected at a much higher velocity
$v_\mathrm{ej}=[GM_\mathrm{tot}/a]^{1/2}$, where $M_\mathrm{tot}$ is
the total mass of the binary.  The hardening time scale was quantified
for a fixed stellar distribution by \citet{1996NewA....1...35Q}, who
found a time scale of
\begin{equation}
t_h(a)\equiv\left|\frac{a}{\dot{a}}\right|=\frac{\sigma}{G\rho aH},
\label{eqn:th_quinlan}
\end{equation}
where $a$ is the binary separation, $\rho$ and $\sigma$ are the
density and one-dimensional velocity dispersion of the stellar
background, and $H$ is a dimensional parameter whose value was found
from scattering experiments to be 16 for a hard, equal-mass binary.
In practice, however, the above expression for $t_h$ is valid only
during the initial stages of the binary's evolution.  As the binary
shrinks further, it ejects stellar mass from the central regions and
modifies the stellar density $\rho$ that appears in equation
(\ref{eqn:th_quinlan}).  This feedback can be quantified using a
simple analytical model given by \citet{2000ASPC..197..221M}, in which
the binary evolution is described by two coupled equations, the first
describing the binary's hardening due to the presence of stars,
\begin{equation}
\frac{d}{dt}\left(\frac{1}{a}\right)=H\frac{G\rho}{\sigma},
\end{equation}
and the second describing the change in stellar density due to
ejection of mass by the hard SMBH binary,
\begin{equation}
\frac{dM_\mathrm{ej}}{d\ln(1/a)}=JM_\mathrm{tot},
\end{equation}
where $M_\mathrm{ej}$ is the ejected mass, and $J$ is another
dimensionless parameter that was measured by
\citet{1996NewA....1...35Q} to be close to unity and nearly independent
of $a$.

By assuming a singular isothermal sphere profile for the stellar
density and assuming that the ejected stellar mass causes a
constant-density core to form at the center of this profile,
\citet{2000ASPC..197..221M} finds that evolution of the binary
separation can be described as
\begin{equation}
\frac{t-t_\mathrm{init}}{t_0}=\frac{a_h}{a}\left[\ln^2\left(\frac{a_h}{a}\right)-2\ln\left(\frac{a_h}{a}\right)+2\left(1-\frac{a}{a_h}\right),\right]
\label{eqn:binary_evol}
\end{equation}
where $a_h$ is as defined in Equation (\ref{eqn:ahard}),
$a(t_\mathrm{init})=a_h$, and $t_0$ is given by
\begin{equation}
t_0=\frac{9\pi
  J^2}{H}\left(\frac{M_\mathrm{tot}}{2m_2}\right)\left(\frac{GM_\mathrm{tot}}{\sigma^3}\right).
\end{equation}
This result is found to closely match with the evolution observed in
N-body simulation.

On the other hand, the timescale for emission of gravitational waves
is given by
\begin{equation}
t_\mathrm{gr}=\frac{5}{256}\frac{c^5a^4}{G^3m_1m_2M_\mathrm{tot}}.
\end{equation}
As a result, the binary will continue to harden only up to the time
when hardening time $t_h=t_\mathrm{gr}$, after which it will coalesce
rapidly due to gravitational wave emission.  Using equation
(\ref{eqn:binary_evol}), it can be shown that this occurs when
$a=a_\mathrm{gr}$ where \citep{2000ASPC..197..221M},
\begin{equation}
\frac{a_\mathrm{gr}}{a_h}\approx A|\ln A|^{0.4},
\end{equation}
and
\begin{equation}
A=9.85\left(\frac{m_1}{m_2}\right)^{0.2}\left(\frac{M_\mathrm{tot}}{2m_2}\right)^{0.4}\left(\frac{\sigma}{c}\right).
\end{equation}
Here $m_1$ and $m_2$ are masses of the components of the SMBH binary.
Finally, we can again use equation (\ref{eqn:binary_evol}) to
calculate the time it takes for the binary to shrink from $a=a_h$ to
$a=a_\mathrm{gr}$ \citep{2000ASPC..197..221M}:
\begin{equation}
t_\mathrm{coal}\approx 8t_0A^{-1}|\ln A|^{8/5},
\end{equation}
which can be simplified as 
\begin{equation}
t_\mathrm{coal}\approx 1.4\times 10^{10} \mathrm{yr}
\left(\frac{m_2}{m_1}\right)^{0.2}\left(\frac{M_\mathrm{tot}}{2m_2}\right)^{0.6}\left(\frac{M_\mathrm{tot}}{10^9
  \mathrm{M}_\odot}\right)\left(\frac{\sigma}{200
  \mathrm{km}/\mathrm{s}}\right)^{-4}.
\label{eqn:tcoal}
\end{equation}

Clearly, there is a possibility for the formation of multiple-SMBH
system if $t_\mathrm{in}<t_\mathrm{coal}$.  These two time scales are
compared in Figure \ref{fig:tcoal} for a halo that has a mass of
$M_0=10^{12}$ M$_\odot$ at $z=0$.  For simplicity, we have fixed the
mass ratio of merging haloes to be $\xi=0.4$.  At each redshift, we
calculate $t_\mathrm{in}$ from equation (\ref{eqn:tin}).  In order to
estimate $t_\mathrm{coal}$ at a given redshift using equation
(\ref{eqn:tcoal}), we first infer the mass of the halo at that
redshift from the fitting function for the halo's assembly history
from equation (\ref{eqn:halo_assembly}).  We then assume that a galaxy
belonging to a satellite halo with mass ratio $\xi$ has merged with
this host halo at this redshift.

In order to estimate the mass of black holes in the nuclei of these
galaxies, we follow the approach of \citet{2007MNRAS.377..957H} in
employing the $M_{\rm bh}$--$\sigma$ relation.  The virial velocity
(defined as the circular velocity at virial radius) for a halo of mass
$M$ at redshift $z$ is given by
\begin{equation}
v_\mathrm{vir}=23.4\left(\frac{M}{10^8h^{-1}M_\odot}\right)^{1/3}\left[\frac{\Omega_m}{\Omega_m^z}\frac{\Delta_c}{18\pi^2}\right]^{1/6}\left(\frac{1+z}{10}\right)^{1/2} \mathrm{km}/\mathrm{s},
\end{equation}
where 
\begin{equation}
\Omega_m^z=\frac{\Omega_m(1+z)^3}{\Omega_m(1+z)^3+\Omega_\Lambda+\Omega_k(1+z)^2},
\end{equation}
and $\Delta_c$ is the overdensity of the halo relative to the critical
density, given for the $\Lambda$CDM cosmology by
\begin{equation}
\Delta_c=18\pi^2+82d-39d^2,
\end{equation}
where $d=\Omega_m^z-1$ \citep{2001PhR...349..125B}.  Further, we
equate the halo virial velocity with the circular velocity $v_c$ of
its constituent spheroid and obtain the velocity dispersion of the
spheroid using the relation \citep{2002ApJ...578...90F}
\begin{equation}
v_c\approx 314\left[\frac{\sigma}{208 \mathrm{km}/\mathrm{s}}\right]^{0.84}\mathrm{km}/\mathrm{s}.
\label{eqn:sphsigma}
\end{equation}
This combined with the $M_{\rm bh}$--$\sigma$ relation
\citep{2002ApJ...574..740T}
\begin{equation}
\frac{\sigma}{208\mathrm{km}/\mathrm{s}}\approx\frac{M_\mathrm{bh}}{1.56\times 10^8\mathrm{M}_\odot}^{1/4.02},
\label{eqn:msigma}
\end{equation}
gives 
\begin{equation}
\left(\frac{M_\mathrm{halo}}{10^{12}\mathrm{M}_\odot}\right)=8.28\left(\frac{M_\mathrm{bh}}{10^8\mathrm{M}_\odot}\right)\left[\frac{\Omega_m}{\Omega_m^z}\frac{\Delta_c}{18\pi^2}\right]^{-1/2}(1+z)^{-3/2}.
\label{eqn:bhmass}
\end{equation}

We obtain the black hole masses in the host and the satellite haloes
using equation (\ref{eqn:bhmass}) and use the spheroid velocity
dispersion from equation (\ref{eqn:sphsigma}) to estimate the
coalescence time from equation (\ref{eqn:tcoal}).  The result is shown
by the dashed line in Figure \ref{fig:tcoal}.

At high redshift, early on in the assembly history of a halo, the
galaxy merger rate is higher than the SMBH binary coalescence rate and
systems with multiple SMBHs can form.  Note that the time scale
$t_\mathrm{coal}$ obtained above will change if effect of loss-cone
replenishment and gas are taken into account.  However,
\citet{2002MNRAS.331..935Y} finds that in realistic spheroidal
galaxies, even loss-cone replenishment is insufficient to cause early
coalescence.

\section{Evolution of multiple SMBHs}
\label{sec:evolution_analytic}

We have described the literature on systems with more than two SMBHs
in \S \ref{prev}.  If the infalling SMBH is less masssive than
either of the components of a pre-existing binary then we expect the
ultimate outcome to be ejection of the smaller SMBH and recoil of the
binary.  \citet{2007MNRAS.377..957H} studied the statistics of close
triple SMBH encounters in galactic nuclei by computing a series of
three-body orbits with physically motivated initial conditions
appropriate for giant elliptical galaxies.  Their simulations included
a smooth background potential consisting of a stellar bulge and a dark
matter halo, and also accounted for the effect of dynamical friction
due to stars and dark matter.  They found that in most cases the
intruder helped the binary SMBH to coalesce via the Kozai-Lidov
mechanism and by scattering stars into the binary's loss cone.  In
this case, the intruder itself was left wandering in the galactic
halo, or even kicked out of the galaxy altogether.  It was also found
that escape of all three black holes is exceedingly rare.

Dynamical evolution of multiple massive black holes in globular
clusters has received much attention \citep{1993Natur.364..421K,
1993Natur.364..423S}.  From these studies, it is expected that systems
with more than two SMBHs will last for about a crossing time.

\section{Simulations}
\label{sec:simulations}

\begin{figure}
\label{fig:prog}
\begin{center}
\includegraphics[scale=0.6]{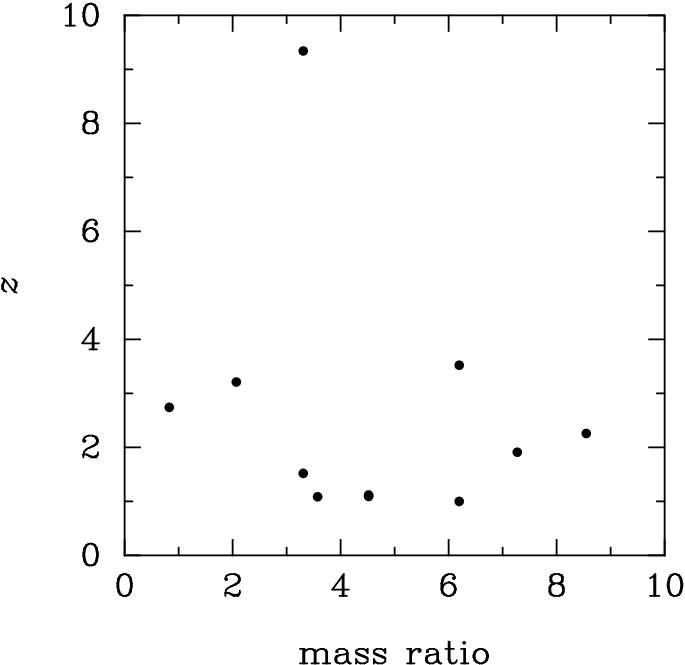} 
\end{center}
\caption{An example merger tree form the Millennium simulation of a
  halo that has a mass of $M_0 \sim 10^{12}$ M$_\odot$.  This
  plot shows major mergers (mass ratio $> 0.1$) \emph{in all branches}
  of the halo's merger tree.}
\end{figure}

\begin{table*}
\begin{center}
\begin{tabular}{c c c c c}
\hline
Simulation & Mass of halo at $z=0$ (M$_\odot$) & Max.\ BH no.\ & SMBH Coalescences & SMBH Escapes \\
\hline
L1 & $1.21\times10^{14}$ & 4 & 7 & 2 \\
L2 & $1.31\times10^{14}$ & 2 & 1 & 1 \\
L3 & $1.31\times10^{14}$ & 2 & 3 & 2 \\
L4 & $1.24\times10^{14}$ & 2 & 5 & 5 \\
L5 & $1.28\times10^{14}$ & 5 & 8 & 4 \\
L6 & $1.31\times10^{14}$ & 6 & 6 & 0 \\
L7 & $1.23\times10^{14}$ & 3 & 2 & 0 \\
L8 & $1.31\times10^{14}$ & 2 & 3 & 1 \\
\hline
\end{tabular}
\end{center}
\caption{Summary of simulations and results for haloes that have a
  mass of $M_0 \sim 10^{14}$ M$_\odot$. The maximum BH number denotes
  the number of black holes in the biggest BH group found in a
  simulation.  The last two columns show number of BH coalescences and
  escapes in the simulation.  A halo with $M_0=10^{14}$ M$_\odot$ has
  average mass $M_{z=6}=5\times 10^{11}$ M$_\odot$.}
\label{table:sims_sdss}
\end{table*}

\begin{table*}
\begin{center}
\begin{tabular}{c c c c c}
\hline Simulation & Mass of halo at $z=0$ (M$_\odot$) & Max.\ BH
no.\ & SMBH Coalescences & SMBH Escapes \\ 
\hline H1 & $1.25\times10^{15}$ & 6 & 4 & 3 \\ 
H2 & $1.65\times10^{15}$ & 2 & 1 & 1 \\ 
H3 & $1.81\times10^{15}$ & 3 & 2 & 0 \\ 
H4 & $1.24\times10^{15}$ & 5 & 6 & 3 \\ 
H5 & $1.37\times10^{15}$ & 3 & 7 & 1 \\ 
H6 & $1.40\times10^{15}$ & 4 & 3 & 0 \\ 
H7 & $1.41\times10^{15}$ & 6 & 9 & 1 \\ 
H8 & $1.45\times10^{15}$ & 3 & 4 & 1 \\ 
H9 & $1.46\times10^{15}$ & 2 & 2 & 0 \\ 
H10 & $1.48\times10^{15}$ & 4 & 7 & 1 \\ 
H11 & $1.54\times10^{15}$ & 2 & 1 & 1 \\ 
H12 & $1.59\times10^{15}$ & 5 & 10 & 1 \\ 
H13 & $1.66\times10^{15}$ & 8 & 15 & 4 \\ 
H14 & $1.71\times10^{15}$ & 4 & 3 & 0 \\ 
H15 & $1.81\times10^{15}$ & 4 & 20 & 7 \\ 
H16 & $1.86\times10^{15}$ & 3 & 7 & 4 \\ 
H17 & $4.04\times10^{15}$ & 8 & 11 & 2 \\ 
\hline
\end{tabular}
\end{center}
\caption{Summary of simulation runs with haloes that have mass $M_0
  \gtrsim 10^{15}$ M$_\odot$ at $z=0$.  Various columns are same as
  Table \ref{table:sims_sdss}.  A halo with $M_0=10^{15}$ M$_\odot$ has
  average mass $M_{z=6}=5\times 10^{12}$ M$_\odot$.}
\label{table:sims_vhmass}
\end{table*}

\begin{figure*}
\begin{center}
\begin{tabular}{ccc}
\includegraphics[scale=0.6]{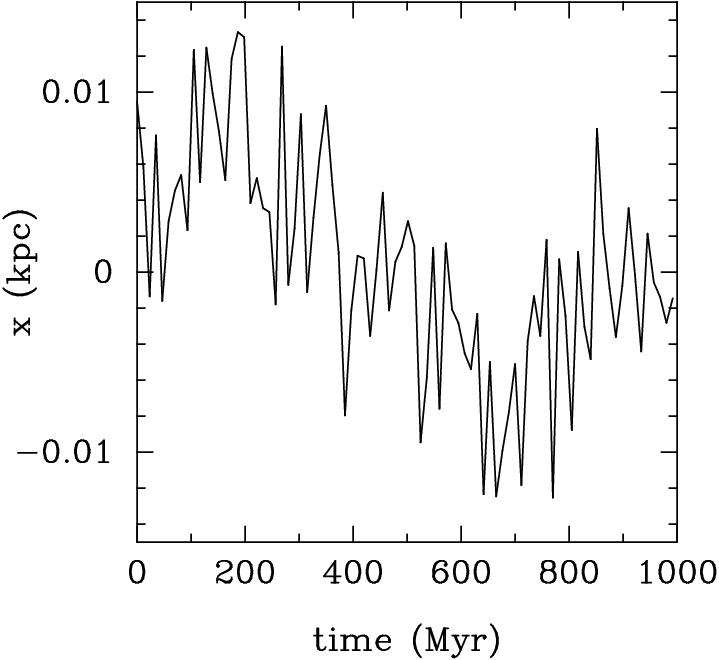} & 
\qquad\qquad &
\includegraphics[scale=0.6]{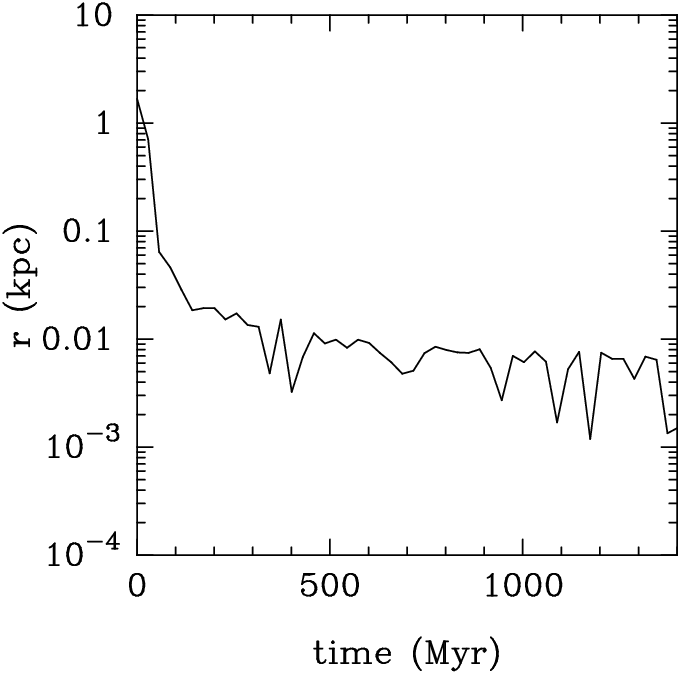} \\
\end{tabular}
\caption{Evolution of single and binary SMBHs in our simulations.  (a)
  The left hand panel shows evolution of the $x$-component of the
  position of a $9.95\times 10^5$ M$_\odot$ back hole near the centre
  of a Hernquist bulge of mass $5.41\times 10^7$ M$_\odot$ and scale
  length of $0.2$ kpc.  The particle mass is $5.411\times10^3$
  M$_\odot$. The secular motion is due to that of the cusp.  (b) The
  right hand panel shows evolution of the separation between SMBHs in
  a binary with initial separation $2$ kpc and eccentricity $0.5$.
  The black hole masses were $8.65\times 10^4$ M$_\odot$ and the
  binary evolved near the center of a Hernquist halo with mass
  $5.41\times 10^7$ M$_\odot$ and scale length of $10.0$ kpc.  The
  particle mass is $5.411\times10^3$ M$_\odot$.}
\end{center}
\label{fig:sample_evolution}
\end{figure*}

In order to accurately calculate the formation and evolution of
galactic nuclei with multiple black holes, we perform direct-summation
N-body simulations of galactic nuclei merging in a cosmological
context.  This essentially involves generating physically consistent
initial conditions for galactic nuclei with SMBHs at high redshift and
evolving them while taking into account the mergers of such nuclei and
the resultant close interaction of their SMBHs.

We obtain merger histories of galactic nuclei by extracting merger
trees of gravitationally bound subhaloes from the Millennium
Simulation
Database\footnote{http://www.mpa-garching.mpg.de/millennium/}, which
stores results of the Millennium Simulation
\citep{2005Natur.435..629S}.  The Millennium Simulation is a pure dark
matter simulation with a $\Lambda$CDM model with $2160^3$ particles in
a periodic cube $500$ h$^{-1}$Mpc on a side.  This corresponds to a
particle mass of $8.6\times 10^8$ h$^{-1}$ M$_\odot$.  The output of
this simulation is stored in $64$ snapshots between $z=127$ and $z=0$.
Particles in each snapshot are grouped into friends-of-friends (FOF)
clusters that are expected to correspond to virialised structures.
Each FOF halo contains substructure of gravitationally bound subhaloes
that can be related to each other across snapshots as progenitors and
descendants.  Because a halo can contain multiple galaxies, we expect
the subhalo merger tree to reflect the merger history of the galaxies
within a halo.  Since the goal of this paper is to understand
formation and evolution of systems of multiple black holes due to the
hierarchical merger history of a galaxy, we extract subhalo merger
trees from the Millennium Simulation Database.  Each such merger tree
typically shows growth of a subhalo via accretion of dark matter
particles and via mergers.  We process these merger trees to keep only
major mergers, which we define to be mergers having mass ratio larger
than $0.1$.  To identify the mass ratio of two subhaloes, we use the
masses of the distinct FOF haloes that these subhaloes were a part of
before the FOF haloes merged.  This is to account for the mass loss of
the satellite subhalo due to tidal stripping after it enters the FOF
group of the host subhalo, but before the eventual merger of the two
subhaloes.  (See discussion in \S 5 of \citealt{2007ApJ...665L...5B}.)
Figure \ref{fig:prog} shows the resultant merger history of a Milky
Way sized halo.  The main reason behind removing minor mergers from
our calculation is that for such mergers the dynamical friction time
taken by the satellite halo to reach the center of the host halo is
longer than the Hubble time.  As a result, in such mergers, we do not
expect the constituent galactic nuclei of these haloes to interact
closely.  Since, as we describe below, we model only the spheroidal
galactic nuclei in our simulations, we only need to account for
mergers in which such nuclei will closely interact.  This approach is
very similar to that used by \citet{2007ApJ...665..187L}, with the
main difference being our use of direct-summation N-body simulations
instead of SPH simulations.

Once we have a galaxy merger tree, we set up the initial conditions of
our simulation in the ``leaves'' of the tree, that is, in haloes that
do not have a progenitor, and follow the evolution using an N-body
calculation.  The initial conditions of our simulation consist of a
stellar spheroid with a Hernquist density profile,
\begin{equation}
\rho(r)=\frac{M}{2\pi}\frac{a}{r(r+a)^3},
\label{eqn:hern}
\end{equation}
where $M$ is the total mass of the spheroid and the scale length $a$
is related to the half mass radius $r_{1/2}$ of the spheroid by
$a=0.414r_{1/2}$.  Values for the parameters $M$ and $a$ were obtained
from the halo mass as follows \citep{2007MNRAS.377..957H}.  We first
obtain the black hole mass $M_\mathrm{bh}$ from the halo mass
$M_\mathrm{halo}$ using Equation (\ref{eqn:bhmass}).  We then use the
empirical relation between the SMBH mass and the spheroid's virial
mass \citep{1998AJ....115.2285M, 2003ApJ...589L..21M,
  2006ApJ...640..114P} to obtain the latter as
\begin{equation}
M_\mathrm{sph}=4.06\times 10^{10} \mathrm{M}_\odot \left[\frac{M_\mathrm{bh}}{10^8\mathrm{M}_\odot}\right]^{1.04}.
\end{equation}
The virial mass of the spheroid is related to its velocity dispersion
$\sigma_e$ and half light radius $R_e$ by
\begin{equation}
M_\mathrm{sph}=\frac{kR_e\sigma_e^2}{G}.
\end{equation}
We follow \citet{2003ApJ...589L..21M} and set $k=3$ to get an average
ratio of unity between this mass estimate and the dynamically measured
masses of galaxies.  The velocity dispersion in the above equation is
usually measured over either a circular aperture of radius $R_e/8$ or
a linear aperture of length $R_e$.  These two methods are in essential
agreement, as argued by \citet{2002ApJ...574..740T}.  Assuming a
constant mass-to-light ratio for the Hernquist profile, we have
$R_e=1.815a$ and the velocity dispersion at radius $R_e/8$ is given by
\begin{equation}
\sigma_e^2=\frac{0.104GM}{a}.
\end{equation}
This lets us obtain the value of the parameter $M$ of the Hernquist
profile as $M=1.765M_\mathrm{sph}$.  The scale length $a$ is readily
obtained as
\begin{equation}
a=\frac{GM_\mathrm{sph}}{3\kappa_1\sigma_\mathrm{bh}^2},
\end{equation}
where $\sigma_\mathrm{bh}$ is obtained using the $M-\sigma$ relation
of equation (\ref{eqn:msigma}).  Having obtained a density profile for
the bulge, we place a black hole at its center and set the black hole
mass to be ten times that obtained from equation (\ref{eqn:bhmass}).
This factor of ten is introduced to keep the ratio between the black
hole mass and the particle mass high enough
\citep{2001ApJ...563...34M, 1996ApJ...465..527M}.  We confirm that the
radius of influence $r_\mathrm{inf}=Gm_\mathrm{bh}/\sigma^2$ of this
black hole is still much smaller than the $a$.  Velocities of the
stars in the spheroid are then generated from the unique, isotropic
velocity distribution that corresponds to the gravitational potential
of the density profile in Equation (\ref{eqn:hern}) and the SMBH
\citep{2002ApJ...574..740T}.  These initial conditions are then scaled
to standard N-body units of $G=1$, $M=1$ and $E=-0.25$, where $M$ is
the total mass of the system and $E$ is its total energy
\citep{1986LNP...267..233H,2003gnbs.book.....A}.  In these units, in
virial equilibrium, the mean square velocity $\langle v^2\rangle=1/2$
and the system's crossing time is $t_\mathrm{cr}=2\sqrt{2}$,
independent of the number of particles.  The conversion factors from
physical units to these N-body units can be easily obtained via
dimensional analysis.

Note that we ignore presence of gas in this set-up.  Simulations of
binary BHs in gaseous environment have not reached sufficient
resolution to establish the role played by gas in evolution of SMBHs
in galactic nuclei \citep{2005LRR.....8....8M, 2009arXiv0906.4339C}.
Moreover, we expect that at high redshifts, quasar activity triggered
by galaxy mergers could efficiently drive gas away from the shallow
potential wells of the galaxies.
  
To perform the actual dynamical evolution of this system, we use the
direct-summation code NBODY6 written by Sverre Aarseth
\citep{1999PASP..111.1333A, 2003gnbs.book.....A}.  This code has been
well-tested for various applications since around 1992.  Its purpose
is to perform an exact integration, without particle softening, of a
large number of particles.  It integrates equations of motion of
individual particles using a fourth-order Hermite method with block
time steps \citep{1992PASJ...44..141M}.  This integrator is coupled
with the Ahmed-Cohen neighbour scheme \citep{1973JCoPh..12..389A},
which selects a subset of neighbours of a particle whose forces on it
are calculated at a higher time resolution that other, more distant,
particles.  This scheme reduces the computational cost from
$\mathcal{O}(N^2)$ to about $\mathcal{O}(N^{1.6})$.  Close two-body
encounters are treated using the Kustaanheimo-Stiefel (KS)
regularization method that eliminates the $r=0$ singularity in
Newtonian gravity by using a coordinate transformation.  Triples,
quadruples and compact subsystems of up to six particles (called
``chains'') are treated using the chain regularization method
\citep{1990CeMDA..47..375M}.  Details of the various algorithms in
this code and their implementation are given by
\citet{2003gnbs.book.....A}.  In all simulations reported in this
paper, the time-step parameter for irregular force polynomial,
$\eta_I$, and the time-step parameter for regular force polynomial,
$\eta_R$ are set to 0.02.  The energy tolerance is set to $Q_E=4\times
10^{-5}$ and the regularized time-step parameter is set to
$\eta_U=0.2$.

We check the stability of our initial conditions by evolving
standalone realizations of the Hernquist bulge with a central BH and
then traverse the merger tree of a given halo using NBODY6, starting
from the initial conditions as described above.  We scale the physical
time between two successive nodes of the tree to N-body units and run
NBODY6 for that duration.  If a merger happens at a certain node, we
place the two galactic nuclei at a distance of 2 kpc apart and evolve
in an head-on approach.  Although such head-on mergers would be
unlikely, we choose it to reduce the computational time while still
retaining some realism.  When two galaxies, that are in equilibrium
separately, merge we expect some transient response in the resulting
dynamics.  However, as discussed by \citet{2001ApJ...563...34M}, any
such effects in the dynamics of the central regions of the merger
remnant of these galaxies are essentially negligible.

Under these conditions, the component black holes approach after a
merger event and the remnant galactic nucleus is left with two black
holes, which gradually harden due to dynamical friction and three-body
interactions with stars in their vicinity.  Black hole coalescence is
implemented in our simulation by monitoring the separation of hard
black hole binaries.  Once members of a SMBH binary get closer than a
fixed distance $d_\mathrm{crit}$, we replace them with a single black
hole with mass equal to the sum of the masses of component black
holes.  In all the runs reported in this paper, we set
$d_\mathrm{crit}=0.1$ pc.  Note that this is the only mechanism in
which black holes grow in our simulations.  Thus, the initial SMBH
masses are set according to the $M-\sigma$ relation, but the later
growth of these SMBHs occurs only via coalescence.

Recoil due to anisotropic emission of gravitational waves is a natural
consequence of asymmetric merger of black holes, either due to unequal
masses or due to unequal spins \citep{1962PhRv..128.2471P,
  1973ApJ...183..657B}.  Until recently, it was unclear whether this
recoil is large enough to be astrophysicaly relevant.  However, recent
results from numerical relativity have revealed the resultant kick
velocities in a variety of merger configurations
\citep{2005PhRvL..95l1101P, 2006PhRvL..96k1102B}.  When the black hole
spins are aligned with each other and with the orbital spin, these
simulations find revoil velocity of $v_\mathrm{recoil}\lesssim 200$ km
s$^{-1}$ \citep{2006ApJ...653L..93B, 2007PhRvL..98w1101G,
  2007CQGra..24...33H, 2009PhRvD..79f4018L}.  In the absence of spins,
this recoil velocity is only a function of the ratio of black hole
masses.  For random orientations of spins, recoil velocities as high
as 2000 km s$^{-1}$ have been obtained \citep{2007ApJ...659L...5C,
  2007PhRvL..98w1102C}.  \citet{2007ApJ...661L.147B} argue that a
circumbinary gas disk can align the binary spins with the orbital axis
thereby reducing $v_\mathrm{recoil}$ to about 200 km s$^{-1}$.  In our
simulations we assume a constant kick velocity of $200~{\rm
  km~s^{-1}}$, which we impart to the remnant of every unequal-mass
binary SMBH coalescence.

We follow the approach of \citet{1992PASJ...44..141M} and keep the
particle number fixed at $N=10^4$ throughout the simulation.  Thus, at
every merger, we combine particles in each merging galactic nucleus
and double the particle mass.  This lets us keep the particle number
high throughout the merger tree of the halo.  The ratio of black hole
mass to the stellar mass is typically a few hundred, which is also
roughly the ratio of the spheroid's total mass to the black hole's
mass.  These values are comparable to other simulations of this kind
\citep{1996ApJ...465..527M, 2001ApJ...563...34M}.

In summary, the unique features of our simulations are: {\it (i)}
kinematically consistent initial conditions with black holes; {\it
(ii)} calculation of mergers of galactic nuclei in a cosmological
setting using merger trees extracted from cosmological N-body
simulations; {\it (iii)} calculation of merger of galactic nuclei
resulting in a formation of SMBH binaries starting from the results of
each nucleus having evolved in isolation; and {\it (iv)} accurate
calculation of SMBH-star and SMBH-SMBH dynamics throughout the
assembly history of a galactic nucleus and its constituent SMBH with
the effect of gravitational wave recoil taken into account.

\section{Results}
\label{sec:results_numerical}

We perform some basic checks on our code, such as ensuring energy
conservation and stable evolution of equilibrium systems.  In all of
our runs, the relative error in the total energy is maintained at
$|\Delta E/E|<4\times 10^{-5}$.  The treatment of BH-BH and BH-star
interaction is handled by the original \textsc{nbody6} code, and is
expected to be accurate.  One caveat here is that the neighbour
criterion in \textsc{nbody6} for regularization of close particles is
based on inter-particle distance.  As a result, while evolving a set
of particles in the vicinity of a massive BH, the code either selects
a large number of particles for chain regularization, or selects every
close pair of particles for two-body regularization.  This usually
slows down the code.  Indeed, in three of our runs the code run time
exceeded practical constraints because of this effect.  These three
runs are excluded from the results presented below.

\begin{figure}
\begin{center}
\includegraphics[scale=0.6]{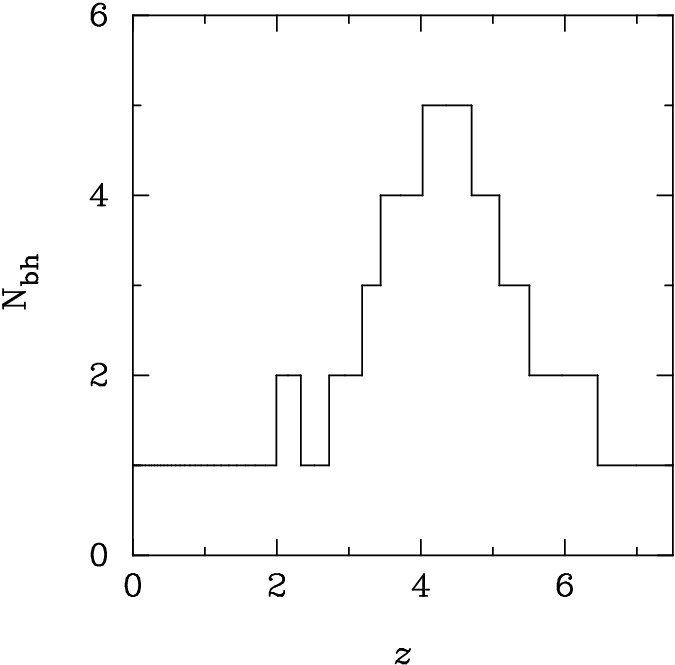}
\end{center}
\caption{Number of black holes as a function of redshift in a
simulation with $M_0=1.29\times 10^{14}$ M$_\odot$.}
\label{fig:nbh_highestmass}
\end{figure}

\begin{figure*}
\begin{center}
\begin{tabular}{ccc}
\includegraphics[scale=0.6]{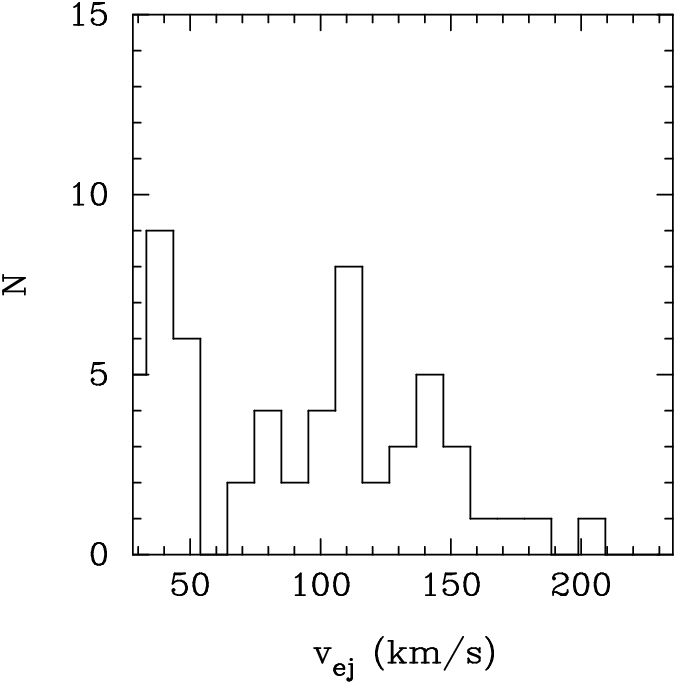} & 
\qquad\qquad &
\includegraphics[scale=0.6]{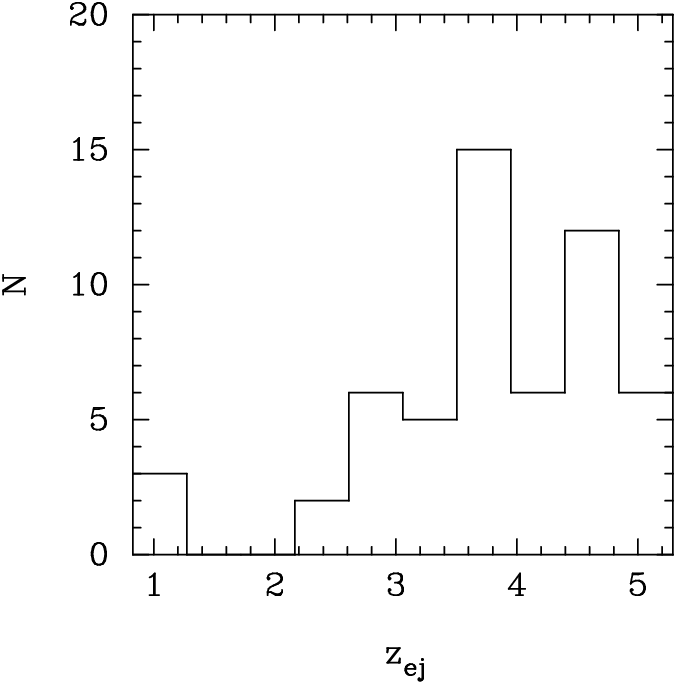} \\
\end{tabular}
\caption{Histograms of ejection velocities of BHs.  Left: Velocities
of ejected black holes in all of our high mass runs.  Note that this
does not include ejected black holes with the highest velocities
($>2000$ km s$^{-1}$).  Right: number of ejections as a function of
redshift in our high mass runs.}
\end{center}
\label{fig:ejection}
\end{figure*}

\subsection{Dynamics of single and binary SMBHs}

In a stellar environment, a single SMBH exhibits a random fluctuating
motion arising due to discrete interactions with individual stars.  As
a result, the effect of the stellar environment on the SMBH can be
decomposed into two distinct components: (1) a smooth component
arising due to the large scale distribution of the whole system, and
(2) a stochastic fluctuating part coming form the interation with
individual stars \citep{2002ApJ...572..371C}.  This random motion is
illustrated in the left hand panel of Figure
\ref{fig:sample_evolution}, which shows evolution of the $x$-component
of the position of a $9.95\times 10^5$ M$_\odot$ back hole near the
centre of a Hernquist bulge of mass $5.41\times 10^7$ M$_\odot$ and
scale length of $0.2$ kpc.  The particle mass is $5.411\times10^3$
M$_\odot$.  As expected, the SMBH wanders around due to stochastic
interactions with the stars in its vicinity.  The mean square
amplitude of these fluctuations is expected to be
\citep{2002ApJ...572..371C, 2003ApJ...596..860M}
\begin{equation}
\langle x^2\rangle\approx \frac{m_*}{m_\mathrm{BH}}r^2_\mathrm{core},
\end{equation}
where $r_\mathrm{core}$ is the radius within which the stellar
distribution flattens out.  The Hernquist distribution that we have
used here does not have a well-defined core, since the density keeps
rising as $r^{-1}$ near the origin.  \citet{2003ApJ...596..860M} argue
that the effective core radius for such distribution can be taken as
the radius of influence of the black hole.  The resultant mean square
value of fluctuations is somewhat smaller that that for Figure
\ref{fig:sample_evolution} by roughly a factor of 2 as is known to
happen in N-body simulations \citep{1997NewA....2..533Q,
2003ApJ...596..860M}.

As described in \S \ref{sec:intro}, the evolution of a binary black
hole in a gas-poor galaxy takes place in three stages.  Right hand
panel of Figure \ref{fig:sample_evolution} shows evolution of the
separation between SMBHs in a binary with initial separation $2$ kpc
and eccentricity $0.5$ in our code.  The black hole masses were
$8.65\times 10^4$ M$_\odot$ and the binary evolved near the center of
a Hernquist halo with mass $5.41\times 10^7$ M$_\odot$ and scale
length of $10.0$ kpc.  The particle mass is $5.411\times10^3$
M$_\odot$.  In the first stage of evolution, the SMBHs sink to the
centre of the galactic nucleus by losing energy via dynamical friction
and become bound to each other.  This stage ends when the separation
between the SMBHs is equal to the radius of influence of the binary
\citep{2005LRR.....8....8M}.  In the second evolutionary stage, the
binary loses energy predominantly ejection of nearby stars via
three-body interaction.  The binary loses energy rapidly in this
stage, which continues until $t\approx 200$ Myr for the case depicted
in Figure \ref{fig:sample_evolution}.  The final stage of the SMBH
binary evolution begins when the rapid hardening of the second stage
stops.  This happens when the binary semi-major axis takes the value
given by Equation (\ref{eqn:ahard}).  The binary semi-major axis is
related to the separation $r$ by
\begin{equation}
\frac{1}{a}=\frac{2}{r}-\frac{v^2}{\mu},
\end{equation}
where $v$ is the relative velocity of the BHs and $\mu$ is the reduced
mass \citep{2004ApJ...602...93M, 2006ApJ...642L..21B,
2007ApJ...671...53M, 2011arXiv1103.0272K}.  In $N$-body simulations,
the last stage is known to have a dependence on the number of
particles $N$ such that the hardening rate decreases with increasing
$N$ \citep{2004ApJ...602...93M}.  For real spherical galaxies, the
binary separation would stop evolving after this point because the
loss cone is empty.

\begin{figure}
\begin{center}
\includegraphics[scale=0.6]{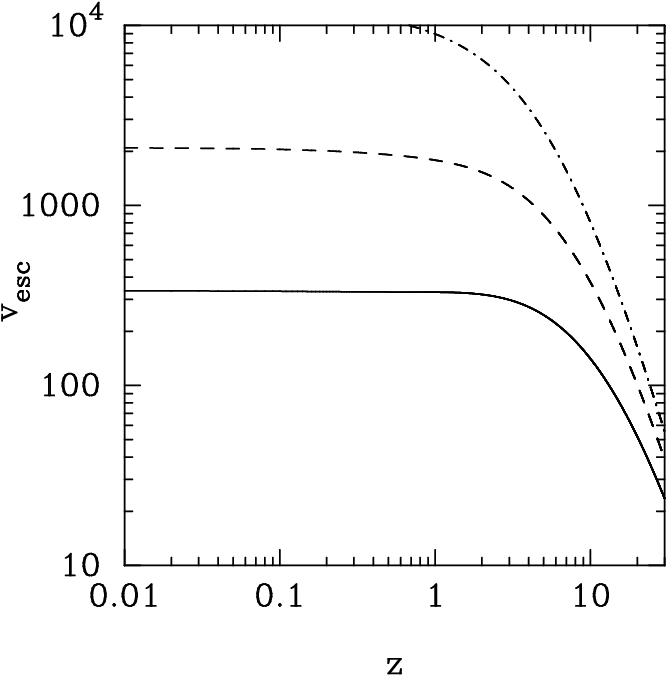} 
\end{center}
\caption{Escape velocities from the bulges of haloes in our three
categories of present-day masses of haloes.  Solid line: $M_0\approx
10^{12}$ M$_\odot$, Dashed line: $M_0\approx 10^{14}$ M$_\odot$,
Dot-dashed line: $M_0\gtrsim 10^{15}$ M$_\odot$.  Note that these are
average values computed from the fitting functions to the Millennium
simulation.  Therefore, case by case comparison with our runs is not
straightforward.}
\label{fig:vesc}
\end{figure}

\subsection{Evolution of nuclei with multiple SMBHs}

\begin{figure*}
\begin{center}
\begin{tabular}{ccc}
\includegraphics[scale=0.5]{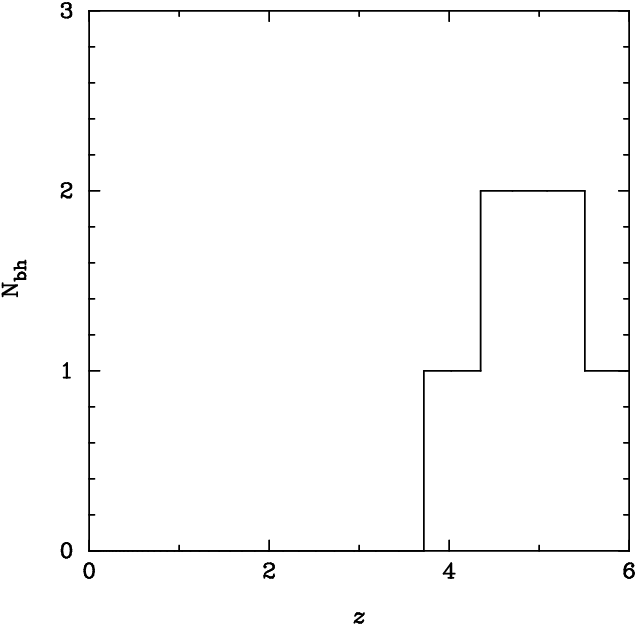} &
\includegraphics[scale=0.5]{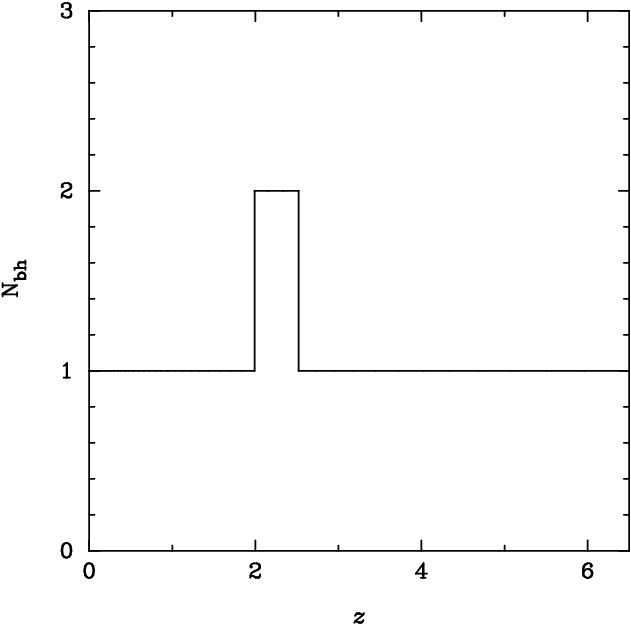} &
\includegraphics[scale=0.5]{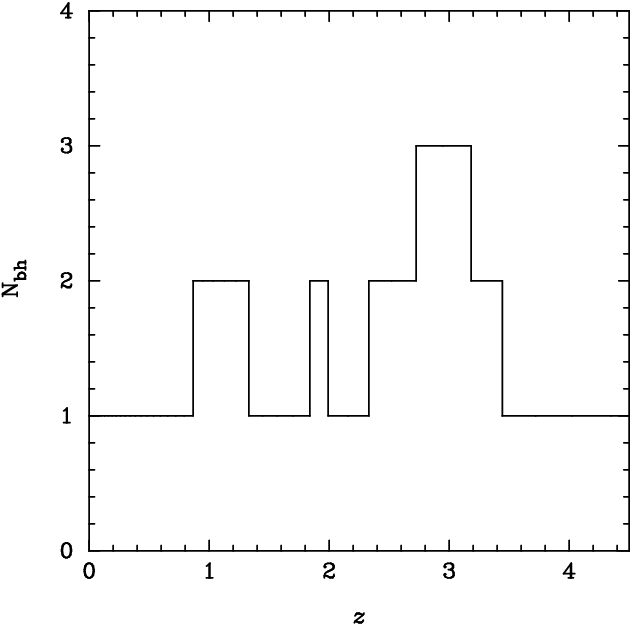} \\
\includegraphics[scale=0.5]{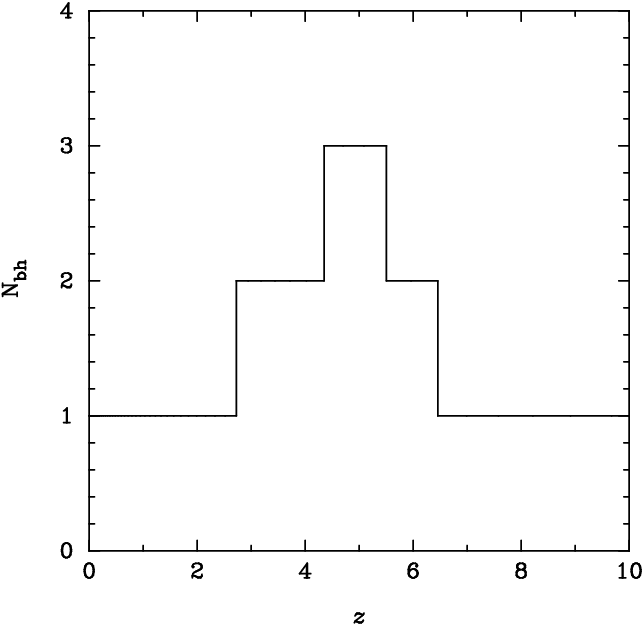} &
\includegraphics[scale=0.5]{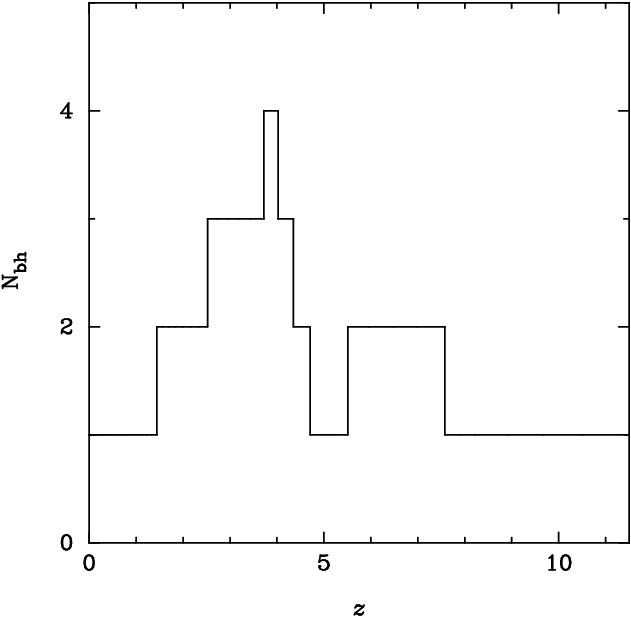} &
\includegraphics[scale=0.5]{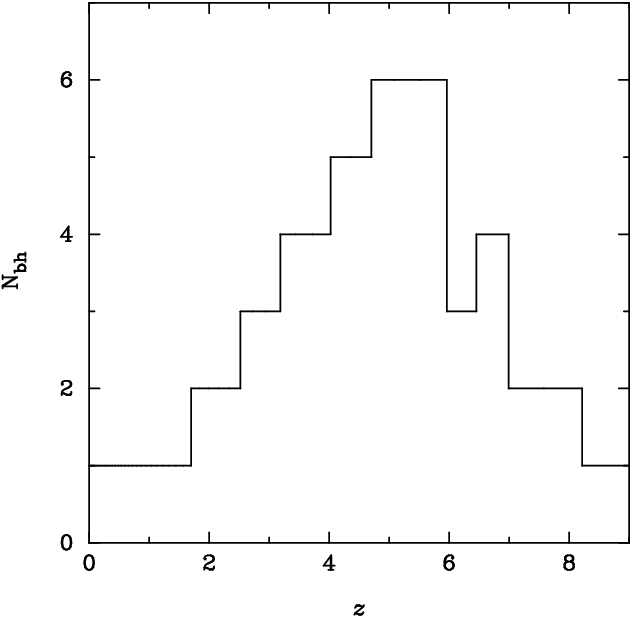} \\
\end{tabular}
\end{center}
\caption{Number of black holes as a function of redshift in a few of our simulation runs.}
\label{fig:runsamples}
\end{figure*}

\begin{figure*}
\begin{center}
\begin{tabular}{ccc}
\includegraphics[scale=0.5]{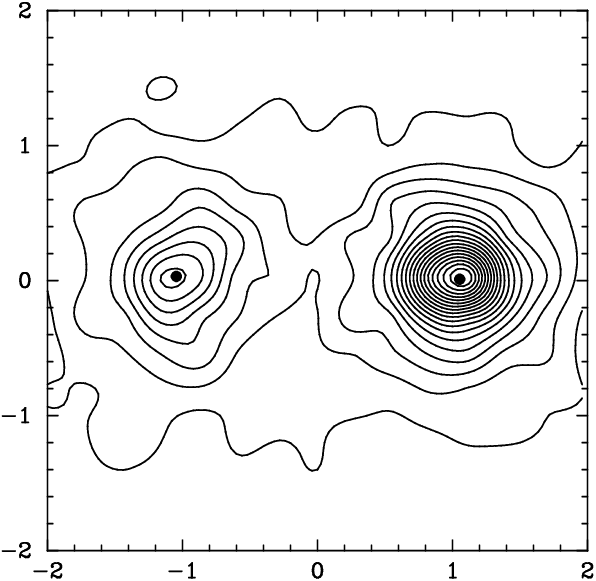} &
\includegraphics[scale=0.5]{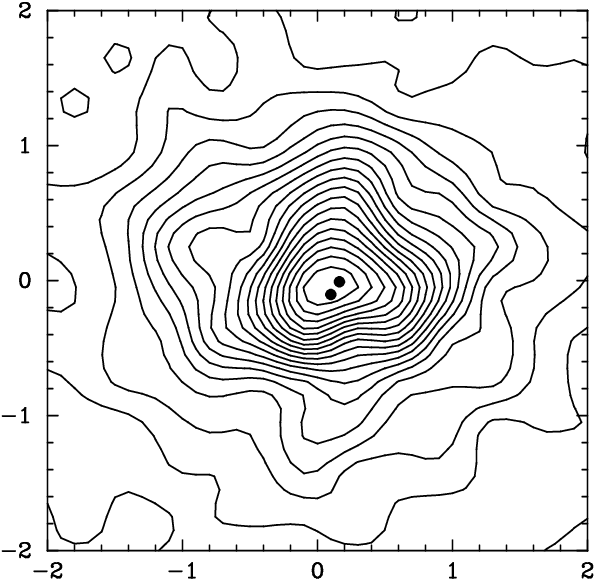} &
\includegraphics[scale=0.5]{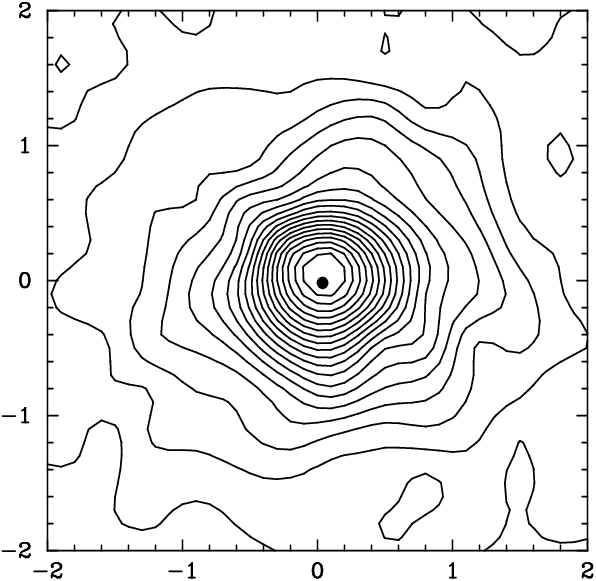} \\ 
\includegraphics[scale=0.5]{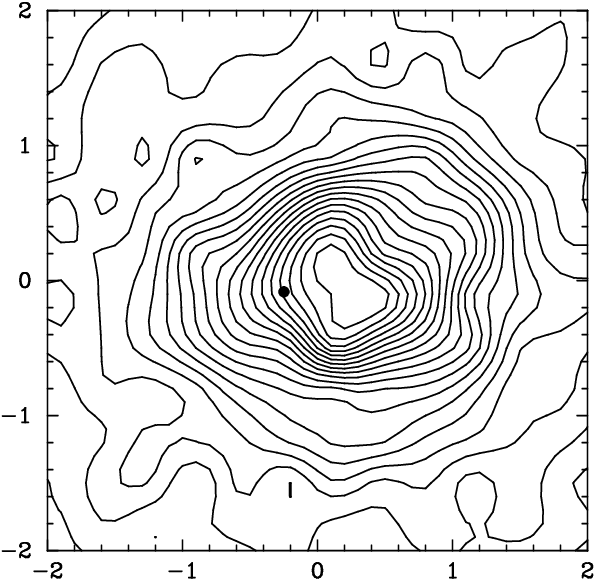} &
\includegraphics[scale=0.5]{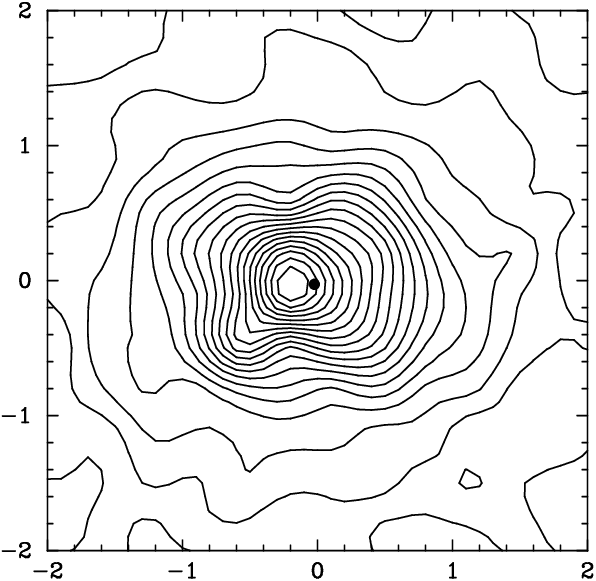} &
\includegraphics[scale=0.5]{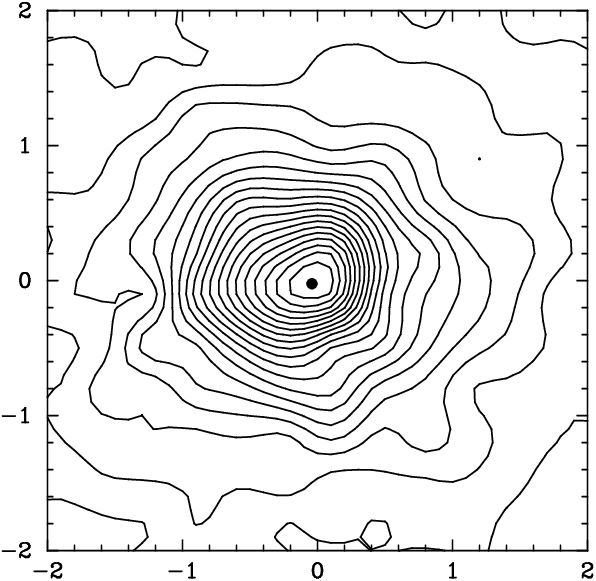} \\
\end{tabular}
\end{center}
\caption{Projected stellar density contours in the presence of a
  binary in the simulation H5.  Each panel is 400 pc on a side.
  Clockwise from top left to bottom right, the redshifts are
  $z=10.073$, $8.54$, $7.27$, $6.19$, $5.28$, and $4.52$.  The total
  time span is about 800 Myr.  Core-SMBH oscillations are clearly
  visible.}
\label{fig:binarysample}
\end{figure*}

\begin{figure*}
\begin{center}
\begin{tabular}{ccc}
\includegraphics[scale=0.5]{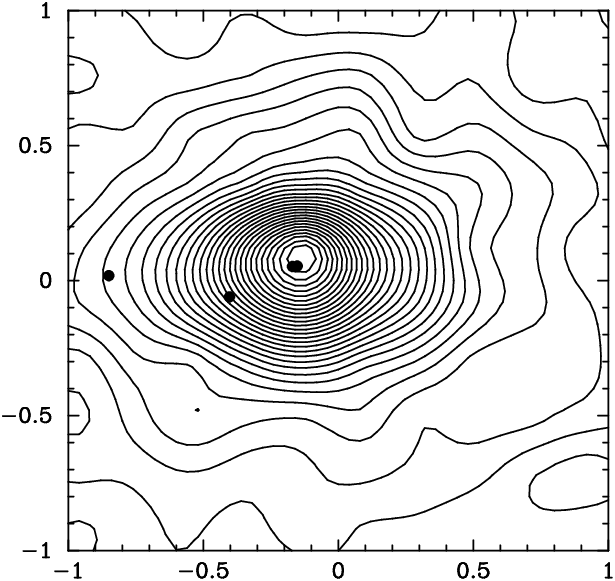} &
\includegraphics[scale=0.5]{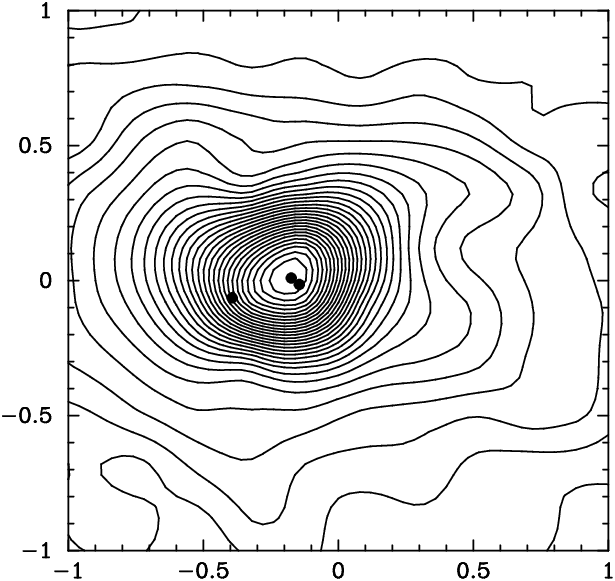} &
\includegraphics[scale=0.5]{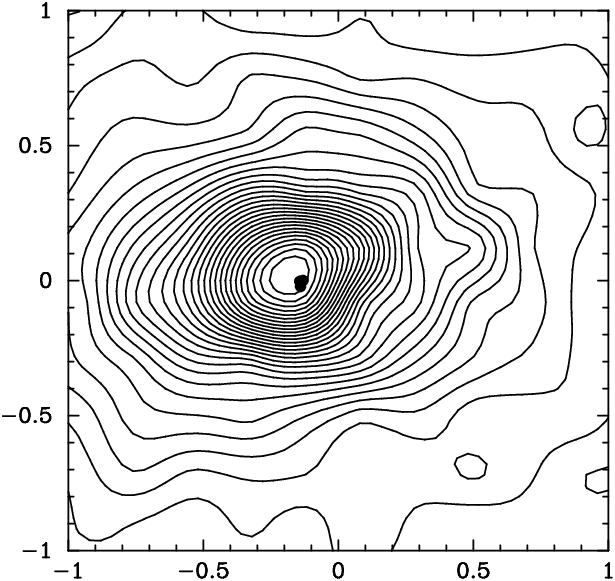} \\
\includegraphics[scale=0.5]{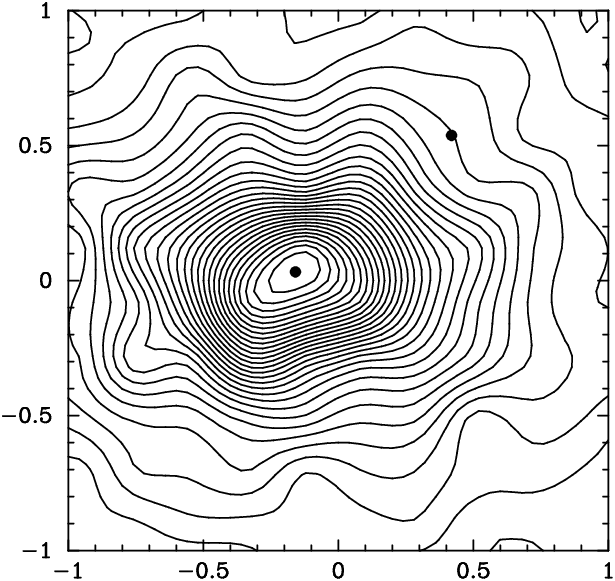} &
\includegraphics[scale=0.5]{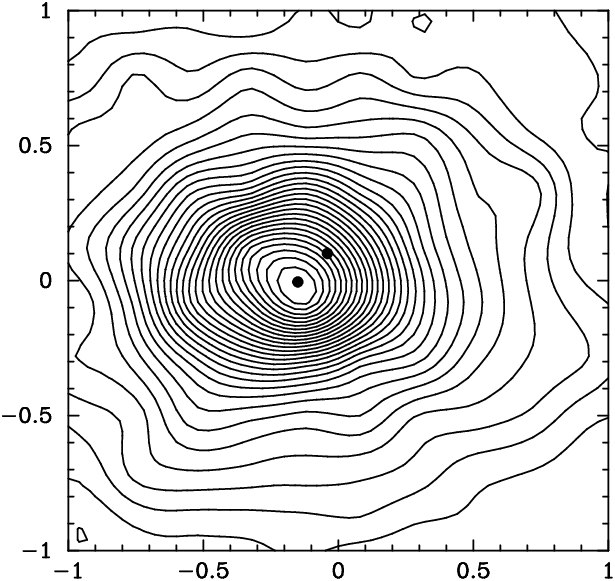} &
\includegraphics[scale=0.5]{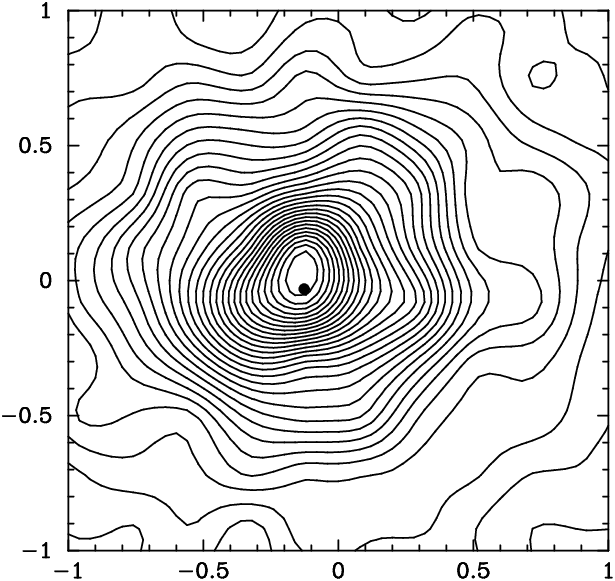} \\
\end{tabular}
\end{center}
\caption{Projected stellar density contours in the presence of
  multiple BHs in the simulation H4.  Each panel is 100 pc on a side.
  The total time span, clockwise from top left to bottom right, is
  about 1 Gyr.  Most BHs are stripped of their cusps in nuclei with
  multiple BHs.}

\label{fig:contour_multiple}
\end{figure*}

\begin{figure*}
\begin{center}
\begin{tabular}{ccc}
\includegraphics[scale=0.5]{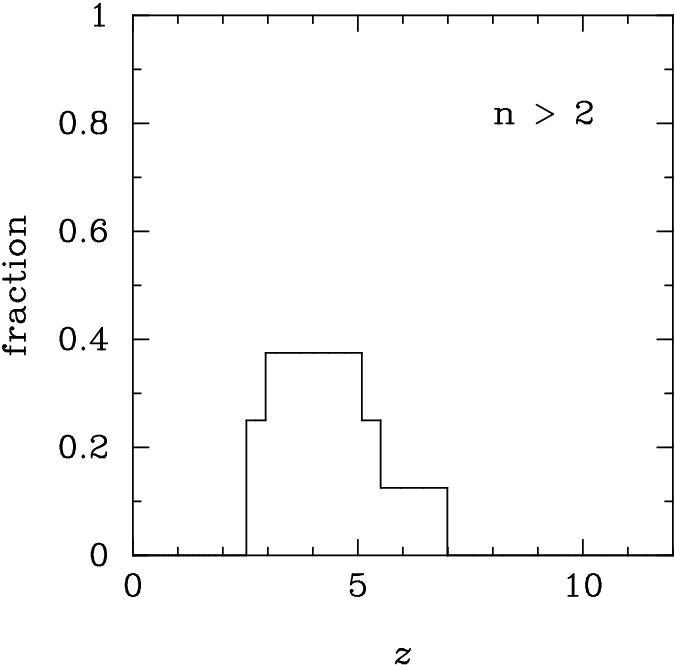} &
\includegraphics[scale=0.5]{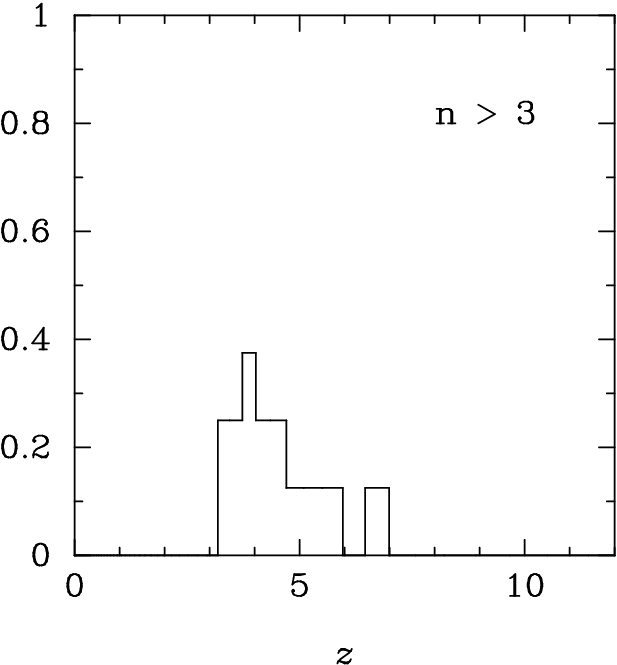} &
\includegraphics[scale=0.5]{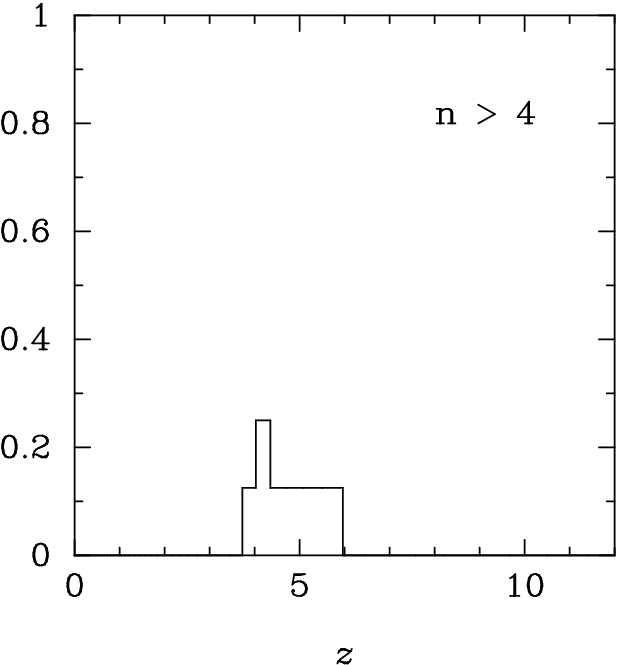} \\
\end{tabular}
\end{center}
\caption{The fraction of runs with multiple SMBHs at different
  redshift bins for haloes with a mass $M_0\sim
  10^{14}\mathrm{M}_\odot$ at $z=0$.  The results of these runs are
  summarised in Table \ref{table:sims_sdss}.  The three panels from
  left to right describe the occurrence of systems with more than 2, 3
  and 4 black holes respectively.  At each redshift, this number can
  be interpreted as the likelihood of finding such systems in haloes
  of mass $M_0\sim 10^{14}\mathrm{M}_\odot$ at $z=0$.  It is seen that
  systems with multiple SMBHs are rare at redshift $z\lesssim 2$.
  Note that a halo with $M_0=10^{14}$ M$_\odot$ will have
  $M_{z=6}=5\times 10^{11}$ M$_\odot$.}
\label{fig:lhood_sdss}
\end{figure*}

\begin{figure*}
\begin{center}
\begin{tabular}{ccc}
\includegraphics[scale=0.5]{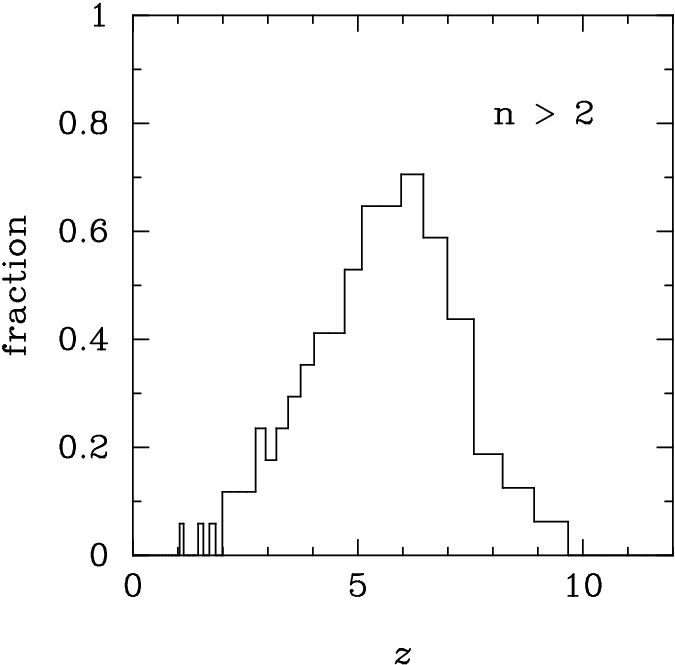} &
\includegraphics[scale=0.5]{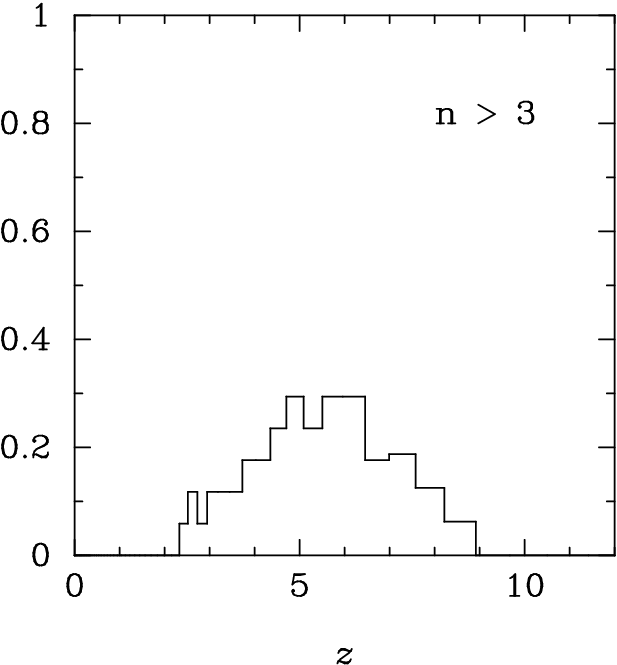} &
\includegraphics[scale=0.5]{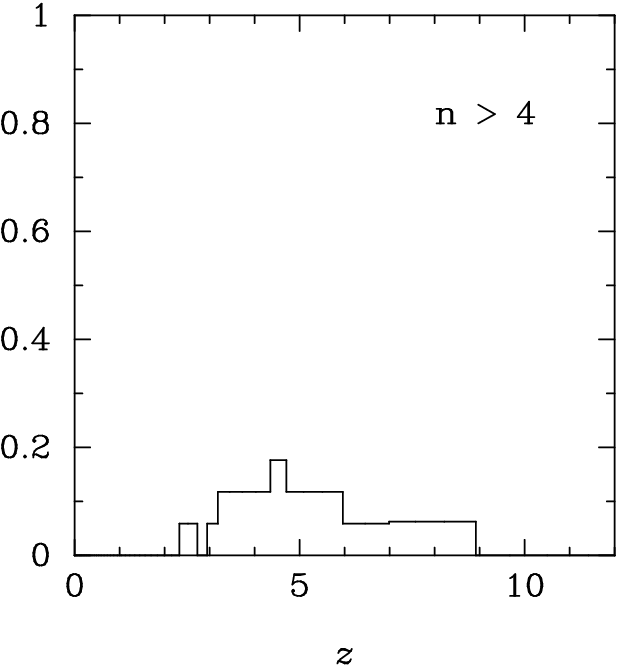} \\
\end{tabular}
\end{center}
\caption{The fraction of runs with multiple SMBHs at different
  redshift bins for halo masses $M_0\gtrsim 10^{15}\mathrm{M}_\odot$ at
  $z=0$.  The results of these runs are summarised in Table
  \ref{table:sims_vhmass}.  The three panels from left to right
  describe the occurrence of systems with more than 2, 3 and 4 black
  holes respectively.  At each redshift, this number can be
  interpreted as the likelihood of finding such systems in haloes of
  mass $M_0\gtrsim 10^{15}\mathrm{M}_\odot$ at $z=0$.  It is seen that
  systems with multiple SMBHs are rare at redshift $z\lesssim
  2$. These results can be compared with those in figure
  \ref{fig:lhood_sdss}.  Nuclei with multiple SMBHs are more likely in
  high mass haloes because of higher merger rate.  Note that a halo
  with $M_0=10^{15}$ M$_\odot$ will have $M_{z=6}\sim 10^{12}$
  M$_\odot$.}
\label{fig:lhood_rare}
\end{figure*}

We now run the simulation along merger trees of haloes drawn from the
Millennium simulation as described in Section \ref{sec:simulations}.
These simulations are described in Tables \ref{table:sims_sdss} and
\ref{table:sims_vhmass}.  We randomly select 8 haloes with mass $M_0$
around $10^{14}$ M$_\odot$ at $z=0$.  These correspond to the typical
haloes ($M\approx M_*$) in the present epoch.  We also randomly select
17 haloes whose present-day mass $M_0$ is in excess of $10^{15}$
M$_\odot$.  These are rare, high mass haloes that are expected to host
the redshift 6 SDSS quasars \citep{2007ApJ...665..187L}.
Additionally, we have also simulated 11 haloes with present-day mass
similar to the Milky Way halo ($M_0 \sim 10^{12}$ M$_\odot$).Using the
prescriptions described in the previous section, and using the N-body
integrator, these simulations tell us about the effect of multiple
mergers of galactic nuclei with SMBHs.

Figure \ref{fig:nbh_highestmass} shows results from a typical
simulation run, for a halo of mass $1.29\times 10^{14}$ M$_\odot$.  We
plot here the number of BHs in the bulge in the main branch of the
galaxy's merger tree at various redshifts.  It is seen that the
central bulge has more than one SMBH for a wide redshift range
($2\lesssim z\lesssim 6$; about $2.5$ Gyr).  For $3\lesssim z\lesssim
5$ (about $1$ Gyr) the bulge holds more than 2 BHs.  The maximum
number of BHs interacting within the bulge in this simulation is 6.
Lastly, the number of BHs reduces to one well before $z=0$ due to
coalescences and ejections.  Note that at the highest redshifts
($z\gtrsim 6$) there are no BHs in the central bulge.  This is simply
an artifact of the limited numerical resolution of the Millennium
simulation, because of which the halo merger tree is not resolved at
these redshifts.  To ensure that this does not affect our results for
$z\lesssim 6$, we set up initial conditions at $z\sim 6$ such that the
BHs are on the $M-\sigma$ relation, and by using a Hernquist bulge
with inner slope $-1$.  In the absence of gas, the systems with
multiple SMBHs form generically, in high mass haloes with frequent of
major mergers.  It is evident than such systems are usually
short-lived and most often these nuclei contain a single SMBH at
$z=0$.  Most SMBHs escape into the halo, where they join a population
of wandering black holes or escape the halo completely.

\begin{figure*}
\begin{center}
\begin{tabular}{ccc}
\includegraphics[scale=0.6]{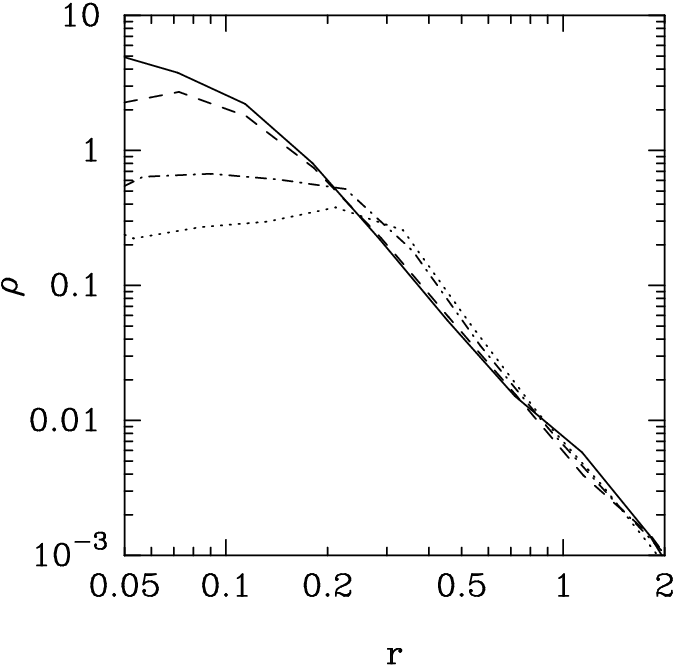} & 
\qquad\qquad &
\includegraphics[scale=0.6]{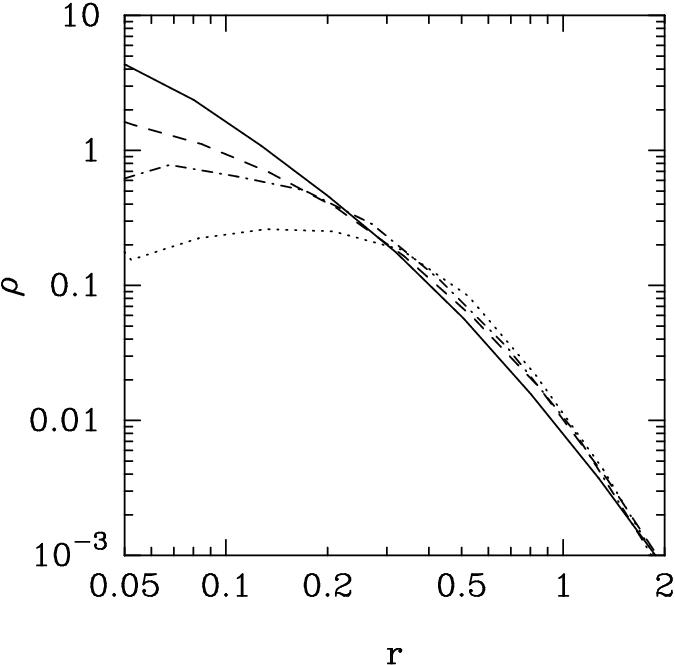} \\
\end{tabular}
\label{fig:core}
\caption{Evolution of density profile for simulations H3 and H5 in
  N-body units.  The solid line is the original Hernquist profile with
  an inner logarithmic slope of $\gamma\approx -1$.  Dashed line shows
  the profile after one SMBH binary coalescence, dot-dashed line after
  the second coalesence and the dotted line after the third
  coalescence.  These plots are shown in N-body units to scale out the
  doubling of the half-mass radius. See text for details.}
\end{center}
\end{figure*}

Similar results from a few other simulation runs for haloes with mass
$M_0 \sim 10^{14}$ M$_\odot$ at $z=0$ are shown in Figure
\ref{fig:runsamples}.  Most of these runs have features similar to the
run described above.  Multiple BH systems form generically and last
for $2-3$ Gyr.  Importantly, most of these galaxies end up with a
single SMBH in their central bulge.  This is in contrast with
expectations from some simple arguments in earlier work
\citep{1992MNRAS.259P..27H}.  About 5\% of galaxies in our simulations
end up with no BHs in their centres at $z=0$.  Tables
\ref{table:sims_sdss} and \ref{table:sims_vhmass} summarize these
features of all our simulations.  The last columns of these tables
show the cumulative number of BHs that were ejected out of the
galactic nucleus throughout the run either due to recoil associated
with emission of gravitational waves or due to many-body interaction
between the BHs.  We find that for most triple and quadruple SMBH
systems in our calculation, gravitational wave recoil is the dominant
mechanism for SMBH escape.  Many-body interaction between SMBHs was
the dominant cause only when the number of black holes was more than
four.  Consequently, for low-mass galaxies in which the number of BHs
is small, almost all escapes were because of gravitational wave
recoil.  Whereas in our low mass galaxy simulations, larger number of
coalescence usually results in large escapers, in the high mass galaxy
simulations, coalescence often does not lead to escape.  In high mass
galaxies, BH-BH interaction is the dominant mechanism behind escaping
SMBHs.  Figure \ref{fig:ejection} summarizes this.  The right hand
panel shows that most ejections happen at high redshifts.  Typical
ejection velocities are seen in the left hand panel.  Ejection
velocities are spread out up to 200 km s$^{-1}$, which is the GW
recoil kick in our simulations.  Note that this plot does not show
kicks with very high velocities, which we describe below.

With the prescription that we have adopted in this paper, we find that
SMBH coalescence happens in each one of our simulations.  Tables
\ref{table:sims_sdss} and \ref{table:sims_vhmass} give the number of
BH coalescences occurring in our simulations.  Due to the limitation
on the particle number, our simulations implement BH coalescence by
replacing a bound binary BH by a single BH whose mass is equal to the
total mass of the binary.  As an example, Figure
\ref{fig:binarysample} shows the merger of two bulges beginning from
initial conditions at redshift 6.7 in the run H5.  In Figure
\ref{fig:binarysample}, the hardening radius is $a_h=0.5$ pc at
$t_h=500$ Myr.  We find the the BHs remain associated with their host
cusps until cusp coalescence.  It is known that by increasing the
effective mass of the BHs, this increases the rate of coalescence of
the BHs by as much as $\sim 6$ times compared to the dynamical
friction time scale.  We also see the homology of density structure
before and after the merger, as reported previously in the literature
\citep{2001ApJ...563...34M}.  However, one prominent difference from
previous works is in the evolution of the density profile in the later
stages of the merger.  In our simulations, each coalescence event is
followed by recoil of the remnant at 200 km s$^{-1}$, which at high
redshift, usually results in the escape of the SMBH from the galaxy.
At relatively low redshifts, the recoiled SMBH returns to the nucleus
in few hundreds Myr.  Because of this recoil, the remnant BH is
detached from its cusp immediately.  At the recoil speed implemented
here, this happens at a much smaller time scale that the local
crossing time scale.  As a result, the only effect of the remnant on
the cusp is due to subsequent core passages.

Usually, most coalescences are assumed to take place due to BH
hardening via BH-star encounters.  In gas-free systems, this leads to
the final parsec problem.  In our simulations, we find that in high
mass haloes, roughly half of the SMBH coalescences are due to
three-body scattering with intruder SMBHs.  This is expected, since in
spite of higher major merger rate, high mass galaxies in our model are
still left with at most two SMBHs at $z=0$. The dominant mechanism of
coalescence is then three body interactions.  Figure
\ref{fig:runsamples} shows an example of the evolution of a multiple
BH system that undergoes three coalescences due to BH-BH dynamics.  We
find violent oscillations of the cusp-BH system as shown in Figure
\ref{fig:binarysample}. This has significant impact on the density
distribution of the core, and also results in off-centre BHs, which
slowly return to the centre of the cusp due to dynamical friction.

About 10\% of SMBH ejections in our simulations occur at very high
speeds of $\gtrsim 2000$ km s$^{-1}$.  In haloes with $M_0\approx
10^{15}$ M$_\odot$ these SMBHs will linger in the outskirts of the
halo for $2-10$ Gyr as can be seen by comparing with the bulge escape
speeds in Figure \ref{fig:vesc}.  The SMBHs in the wandering phase
that are introduced via this mechanism have markedly different
properties than the BHs introduced due to galaxies that have not yet
reached the host galaxy's center so as to have a close encounter
\citep{2003ApJ...582..559V}.  The main difference is that our ejected
black holes are much more massive than those in the other
category. Moreover, the velocity of ejected SMBHs will typically be
higher that black holes in the other category, which have already
experience significant dynamical friction.  Three of the 30 BH
ejections in our runs are ejected binaries.

\subsection{Likelihood of nuclei with multiple SMBHs at high redshift}

From the results of our simulations, we can estimate the likelihood of
galactic nuclei with multiple black holes at high redshifts.  The
histograms in Figures \ref{fig:lhood_sdss} and \ref{fig:lhood_rare}
show fraction of runs with multiple SMBHs at each redshift for haloes
with present-day masses of $\sim 10^{14}$ M$_\odot$ and $\sim 10^{15}$
M$_\odot$, respectively.  The three panels from left to right describe
the occurrence of systems with more than 2, 3 and 4 black holes
respectively.  At each redshift, this number can be interpreted as the
likelihood of occurrence of such systems at that redshift.

Systems with more than 2 SMBHs are generically expected in the central
galaxies of haloes with $M_0\gtrsim 10^{14}$ M$_\odot$ at around
$z\gtrsim 3$.  On the other hand, few galaxies hold multiple black
holes at redshifts $z\lesssim 2$ because the galaxy merger rate is low
at these redshifts and the BHs have sufficient time to coalescence.
This is consistent with the expectation from our heuristic analysis of
Section \ref{sec:formation_analytic}.  In other words, multiple black
hole systems are numerous at around redshifts of 6, when there are
many major mergers in the system.  Our numerical simulations show that
such systems can exist in sufficiently long-lived configurations of
SMBHs separated on pc--kpc scale.  Note that these histograms show the
likelihood of such systems to be zero at redshifts $z\gtrsim 10$.
However, this is simply because the Millennium simulation merger trees
do not resolve progenitors at these redshifts.  As mentioned before,
we have minimized the effect of this shortcoming on our results by
requiring that the SMBHs always follow the $M-\sigma$ relation
initially.

High mass galaxies ($M_0\approx 10^{15}$ M$_\odot$) are more likely to
have multiple BHs in their nuclei at higher redshift.  About 60\% of
these galaxies have more than 2 BHs between redshifts $z\approx 2$ and
10.  This fraction is less than 40\% for the low mass galaxies
($M_0\approx 10^{14}$ M$_\odot$) The likelihood of occurrence of more
than 3 and 4 BHs is similar, about 30\%, in the two categories of
simulation.  However, for the high mass galaxies this likelihood is
spread out over a wider range in redshift, again due to the higher
rate of major mergers.

It is extremely rare for Milky Way-sized galaxies (halo mass
$M_0\approx 10^{12}$ M$_\odot$) to have more than three SMBHs in their
nuclei at any moment in their assembly history.  Indeed, in our
simulations of these galaxies, only one run shows a triple BH system.
The main reason behind this is the smaller number of major mergers for
these galaxies.  Moreover, it is easier for SMBHs to escape the nuclei
of predominantly small mass progenitors of these galaxies.

\subsection{Effects on the stellar distribution}

\begin{figure}
\begin{center}
\includegraphics[scale=0.6]{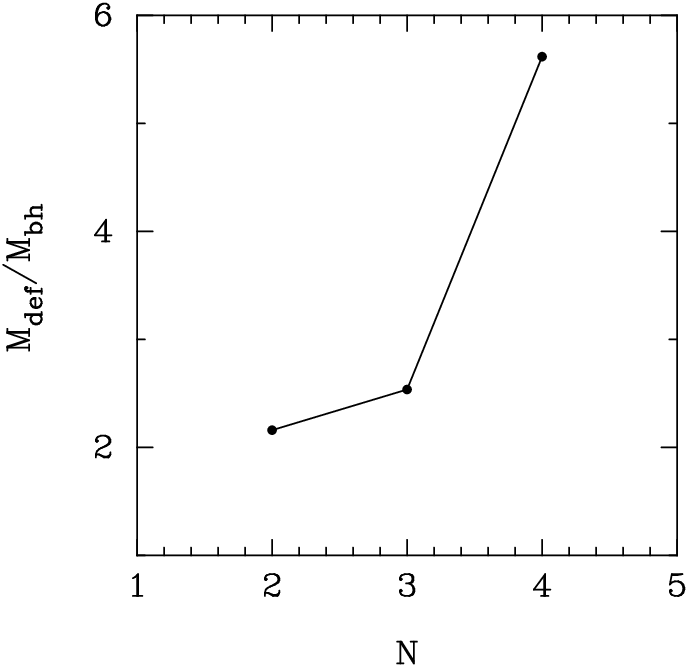} 
\end{center}
\caption{Mass deficiency versus number of coalescences averaged over
ten simulation runs. The presence of multiple SMBHs generally leads to
larger mass deficiency compared to a single hard SMBH binary.}
\label{fig:mdef}
\end{figure}

Most bulges and early-type galaxies have a shallow cusp near their
centre.  The mass distribution in this region can be described as a
power law $\rho\propto r^{-\gamma}$.  Most galaxies have slope
$0.5\lesssim\gamma\lesssim 2.0$ \citep{2006ApJS..164..334F,
  2006ApJ...648..890M}.  We expect the constituent SMBH in the bulge
to affect the mass distribution within its radius of influence.  Only
two galaxies, the Milky Way \citep{2003ApJ...594..812G} and M32
\citep{1998AJ....116.2263L}, have been resolved at these small
distances.  Both these galaxies have $\gamma\approx 1.5$ in their
innermost regions.

It is commonly postulated that cores can form in elliptical galaxies
and spiral bulges due to mass ejection by a hard binary SMBH
(e.g.\ \citealt{2001ApJ...563...34M}).  However, the mass ejected by a
hard binary is of the order of the black hole mass.  In other words,
the mass deficiency $M_\mathrm{def}$, which is the difference between
the mass of the initial and final density distribution in a region
around the centre, is roughly $M_\mathrm{bh}$, the total mass of the
SMBH binary.  The possibility of enhanced mass deficit because of
repeated core passages of recoiled black holes
\citep{2008ApJ...678..780G} and due to repeated mergers
\citep{2006ApJ...648..976M} has been considered in the literature.
Our simulations allow us to understand the effect of both of these
factors in addition to the mass deficit produced by simultaneous
presence of multiple SMBH in the galactic bulge.

Figure \ref{fig:core} shows the cusp evolution in two of our
simulations, each of which has four SMBHs and three coalescences.
Density profiles after each coalescence is shown.  Strong core
formation is clearly seen.  We calculate
$M_\mathrm{def}/M_\mathrm{bh}$ for ten such runs and show the average
result in Figure.  Clearly $M_\mathrm{def}/M_\mathrm{bh}$ is much
larger when multiple SMBHs are present.  Values of
$M_\mathrm{def}/M_\mathrm{bh}\approx 5$ have been observed in large
elliptical galaxies \citep{2004ApJ...613L..33G, 2006ApJS..164..334F,
  2010MNRAS.407..447H}.  Our model explains the occurrence of such
systems.  Since the star-star relaxation time in large elliptical
galaxies in expected to be $\sim 10^{10}$ yr, we can expect them to
carry the signature of core formation at high redshift due to multiple
SMBHs.  At lower redshift our simulation are applicable to spiral
bulges, which have much lower relaxation time scale ($\sim 10^{9}$
yr). Indeed in the runs where a single black hole is left for
$z\lesssim 2$, we find the formation of a Bahcall-Wolf cusp.  This is
consistent with the observed structure of the Milky Way bulge.

The above considerations regarding cores in galaxy luminosity profile
are also applicable to dark matter cores.  The ejection of dark matter
particles by the black holes will produce a core similar in size to
the stellar core.

\section{Observational Signatures}
\label{sec:signatures}

From the results of our simulations described above, we expect about
30\% of the galaxies within haloes with a present-day mass of
$M_0\approx 10^{14}$ M$_\odot$ to contain more than two SMBHs at
redshifts $2\lesssim z\lesssim 6$.  For more massive haloes with
$M_0\gtrsim 10^{15}$ M$_\odot$, this fraction is almost 60\%.
However, since few such systems have been unambiguously observed so
far, we consider some observational signatures that would indicate
their existence\footnote{Some systems with triple active galactic
  nuclei (AGNs) were reported so far.  Examples are NGC 6166 and 7720
  \citep{1984ApJ...279...13T} and SDSSJ1027+1749
  \citep{2011arXiv1104.3391L}.  The first two objects are cD galaxies
  at $z\approx 0.03$ and the latter is at $z\approx 0.06$.  All three
  are kpc-scale triples.  It is possible that NGC 6166 is simply a
  superposition of a central cD galaxy and two low-luminosity
  elliptical galaxies \citep{1998AJ....116.2263L}.}.  Apart from their
effect on the stellar mass distribution, multiple SMBH systems lead to
an enhanced rate of tidal disruption of stars, modified gravitational
wave signals compared to isolated BH binaries, and slingshot ejection
of SMBHs from galaxies at high speeds.

From the results of scattering experiments,
\citet{2009ApJ...697L.149C} found that the stellar tidal disruption
rates due to three-body interactions between a hard, unequal-mass SMBH
binary with fixed separation and a bound stellar cusp is higher by
several orders of magnitude than the corresponding rates for a single
SMBH.  In particular, they find that the stellar tidal disruption rate
is about 1 yr$^{-1}$ for an isothermal stellar cusp with $\sigma=100$
km s$^{-1}$ containing an SMBH binary of total mass $10^7$ M$_\odot$.
In comparison, the corresponding rate for a single $10^7$ M$_\odot$
black hole is about $10^{-4}$ yr$^{-1}$.  The duration of the tidal
disruption phase is about $10^5$ yr.  This enhancement in the tidal
disruption is due to the Kozai-Lidov effect and due to chaotic
resonant scattering \citep{2011ApJ...729...13C}.  Tidal disruption of
a star results in about half of the stellar mass being inserted in
bound elliptical orbits.  When it falls back in the black hole, this
mass gives rise to a bright UV/X-ray emission (``tidal flare'')
lasting for a few years.  One such event may have already been
recently observed in the form of high-energy transients that can be
modeled as sudden accretion events onto an SMBH
\citep{2011arXiv1104.3356L, 2011arXiv1104.3257B, 2011arXiv1106.3568Z}.

We expect similar enhancement in the rate of stellar tidal disruption
in systems with multiple black holes.  Firstly, the presence of
multiple SMBHs increases the combined tidal disruption cross section
of the black holes.  (Although this will only enhance the tidal
disruption rate by a factor of a few.)  Secondly, even before they
closely interact, the presence of a third SMBH affects the tidal
disruption event rate onto an SMBH binary by scattering stars into the
binary's loss cone at a rate that increases as inverse square of its
separation from the binary \citep{2007MNRAS.377..957H}.  Thirdly, as
we saw above, multiple SMBH systems are likely to contain recoiled
black holes, which have been kicked either due to anisotropic
gravitational wave emission after coalescence, or due to the
gravitational slingshot.  Sudden recoil promptly fills the loss cone
of these black holes.  The resultant enhancement in the tidal
disruption event rate can be substantial, increasing it up to 0.1
yr$^{-1}$ \citep{2011MNRAS.412...75S}.  Furthermore, if their recoil
velocity is not too high, these recoiled SMBHs oscillate around the
stellar core with decreasing amplitude due to dynamical friction.
This motion results in their repeated passages through the stellar
core, thereby increasing the stellar tidal disruption event rate.

Another observational signature of systems with multiple SMBHs is
gravitational waves (GWs).  The GW emission from binary and triple
SMBHs has been studied in the literature \citep{2003ApJ...590..691W,
  2004ApJ...611..623S, 2010MNRAS.402.2308A}.  Space-based detectors
like the Laser Interferometer Space Antenna (LISA) are expected to be
sensitive in the frequency range $\sim 10^{-4}$--$10^{-1}$ Hz.  This
corresponds to the inspiral of SMBH systems with total mass $\sim
10^4-10^{10}$ M$_\odot$.  Pulsar timing arrays (PTAs) like the Parkes
PTA \citep{2008AIPC..983..584M} and the European PTA
\citep{2008AIPC..983..633J} and ground-based detectors like the North
American Nanohertz Observatory for Gravitational Waves
\citep{2009arXiv0909.1058J} are sensitive to even lower frequencies of
$\sim 10^{-8}$--$10^{-6}$ Hz.

\citet{2011PhRvD..83d4030Y} studied modifications due to the presence
of a secondary SMBH in the waveform of an extreme mass-ratio inspiral
(EMRI) of a stellar mass objects into an SMBH.  They find that a
$10^6$ M$_\odot$ SMBH will produce detectable modifications if it is
within a few tenths of a parsec from the EMRI system, although this
distance increases for higher mass SMBHs.  In this paper, we have
quantified the presence of such `massive perturbers.'  The resultant
modifications to gravitational waveforms will be a distinct signature
of multiple-SMBH systems.  Futhermore, such systems often contain
binaries that have phases of very high eccentricities, created via
mechanisms like the Kozai-Lidov effect \citep{2007MNRAS.377..957H}.
Such binaries are expected to to emit intense bursts of high-frequency
gravitational waves at the orbital periapsis
\citep{2010MNRAS.402.2308A}.  As a result, sources that would normally
emit outside of the frequency windows of planned gravitational wave
searches may be shifted into observable range.  For example,
\citet{2010MNRAS.402.2308A} find that a few to a hundred gravitational
wave bursts could be produced at a detectable (1 ns) level within the
PTA frequency range if the fraction of SMBH triplets is $\geq 0.1$.

Presence of triple SMBHs also has important implications for
gravitational wave searches using matched-filtering by possibly
requiring additional waveform templates \citep{2011MNRAS.412..551A}.

Lastly, an observable signature of these systems will be the presence
of wandering SMBHs in the large haloes \citep{2007MNRAS.377..957H}.
We have shown that about 10\% of the SMBHs are ejected at velocities
$>2000$ km s $^{-1}$ due to the slingshot mechanism.  This high-speed
black holes will spend $1-10$ Gyr in the outskirts of the halo.
However, it is not clear whether detecting this population of
wandering black holes will be possible.

\section{Conclusions}
\label{sec:conclusions}

In this work, we have addressed the formation of galactic nuclei with
mutiple SMBHs.  We performed accurate N-body simulations of mergers of
galactic nuclei with SMBHs in a cosmological setting.  Our calculation
uniquely incorporated cosmological mergers of galaxies with an
accurate treatment of dynamical interactions between SMBHs and stars,
which we achieved using the direct summation N-body code, NBODY6.  The
need for such simulations has been recognized in the literature
\citep{2005LRR.....8....8M}.  Our main conclusions are as follows:

\begin{itemize}
\item In the absence of gas, high mass galaxies ($M_0 \gtrsim 10^{14}$
  M$_\odot$ at $z=0$) are generically expected to have had multiple
  SMBHs in their nuclei during their assembly history.  Our
  simulations suggest that $\sim 30$\% galaxies within haloes with a
  present-day mass of $M_0\approx 10^{14}$ M$_\odot$ ($M_{z=6}\approx
  10^{11}$ M$_\odot$) contain more than two SMBHs at redshifts
  $2\lesssim z\lesssim 6$.  For more massive haloes, with $M_0\gtrsim
  10^{15}$ M$_\odot$ ($M_{z=6}\approx 10^{12}$ M$_\odot$), this
  fraction is almost 60\%.  This is in contrast to lower-mass galaxies
  ($M_0 \approx 10^{12}$ M$_\odot$; $M_{z=6}\approx 10^{10}$
  M$_\odot$), which rarely host more than two SMBHs in their nuclei at
  any moment in their assembly history.

\item High mass galaxies as well as their low mass counterparts are
  rarely expected to retain more than two SMBHs in their nuclei at the
  present epoch.  SMBH coalescence and ejection reduces the number of
  SMBHs on the time scale of a Gyr.  Furthermore, major mergers are
  rare at lower redshift.  We also find that the number of SMBHs in
  galactic nuclei is rarely reduced to zero at $z=0$.  Less than 5\%
  of our high-mass runs resulted in such galaxies.

\item SMBH coalescence is common at high redshifts. Subsequent recoil
  due to anisotropic gravitational wave emission often results in
  escaping SMBHs.  Some of these SMBHs add to the wandering population
  of black holes in the galactic halo.  In a few cases, this process
  also results in galactic nuclei with no SMBH near their centres.
  BH-BH interaction also leads to ejected SMBHs via the slingshot
  mechanism.  While most of ejected SMBHs have velocities $\lesssim
  500$ km s$^{-1}$, about 10\% SMBHs are ejected at very high
  velocities exceeding 2000 km s$^{-1}$.  We also find binary SMBH
  ejection in $\lesssim 10 \%$ of the cases.

\item Multiple SMBHs have a strong effect on the stellar distribution
  due to three-body interactions and core passages. The resulting mass
  deficit is usually much larger than that due to a single SMBH binary
  because of resonant BH-BH interactions and GW recoil of the BH
  remnant.  We observe long-term oscillations of the BH-core system
  that could explain observations of offset AGNs.  This has
  implications for recent observations by \citet{2010ApJ...717..209C}
  of a $z=0.359$ system that potentially contains a recoiled BH.

\item The presence of multiple SMBHs will have important effects on
  the rate of tidal disruption of stars in galactic nuclei due to
  enhanced tidal disruption cross section, scattering of stars by
  other BHs, prompt loss cone refilling due to GW recoil and
  gravitational slingshot.  Similarly, the presence of more than two
  BHs in a hierarchical triple is expected to leave a signature in the
  GW emission from the inner binary.  This signature could be observed
  with future GW observatories, such as LISA.  Finally, we also expect
  such systems to give rise to a distinct population of wandering
  SMBHs that could travel in large haloes over long time scales of a
  few Gyrs.
\end{itemize}

The presence of gas could alter the above picture to some extent.
However, simulations of binary BHs in gaseous environment have not
reached sufficient resolution to confirm this.  Moreover, we expect
that at high redshifts, AGN activity triggered by galaxy mergers could
efficiently drive gas away from the shallow potential wells of the
galaxy.  Our work can also be extended by calculating late stages of
binary SMBH evolution more consistently.  New regularization
techniques to do this are now available \citep{2003gnbs.book.....A};
we defer their use to future work. Furthermore, multiple SMBH systems
can also form in additional ways, for example by fragmentation of
disks \citep{2004ApJ...608..108G}.  However, these systems would
evolve by migration \citep{2011arXiv1104.2322K} on a much shorter time
scale than considered here.
 
\section*{Acknowledgements}

We acknowledge advice on various aspects of our simulations from
Sverre Aarseth and would like to thank him for making his codes
available.  GK also acknowledges discussion with Jasjeet S. Bagla,
Hagai Perets and Yue Shen, and the Institute for Theory and
Computation for hospitality.  Our simulations were run on the Odyssey
cluster supported by the FAS Sciences Division Research Computing
Group at Harvard University.  The Millennium Simulation databases used
in this paper and the web application providing online access to them
were constructed as part of the activities of the German Astrophysical
Virtual Observatory.  This research was supported in part by a
Fulbright-Nehru Professional and Pre-doctoral Fellowship from the
US-India Educational Foundation, by NSF grant AST-0907890 and NASA
grants NNX08AL43G and NNA09DB30A.

\bibliographystyle{mn2e}
\bibliography{refs}
\end{document}